\pdfoutput=1
\RequirePackage{ifpdf}
\ifpdf 
\documentclass[pdftex]{sigma}
\else
\documentclass{sigma}
\fi

\usepackage{euscript}
\usepackage[all,2cell]{xy}
\UseTwocells

\numberwithin{equation}{section}

\begin{document}

\allowdisplaybreaks

\renewcommand{\PaperNumber}{024}

\FirstPageHeading

\ShortArticleName{M-Theory with Framed Corners and Tertiary Index Invariants}

\ArticleName{M-Theory with Framed Corners\\
and Tertiary Index Invariants}

\Author{Hisham SATI}

\AuthorNameForHeading{H.~Sati}

\Address{Department of Mathematics, University of Pittsburgh, Pittsburgh, PA 15260, USA}
\Email{\href{mailto:hsati@pitt.edu}{hsati@pitt.edu}}

\ArticleDates{Received March 19, 2013, in f\/inal form March 01, 2014; Published online March 14, 2014}

\Abstract{The study of the partition function in M-theory involves the use of index theory on a~twelve-dimensional
bounding manifold.
In eleven dimensions, viewed as a~boundary, this is given by secondary index invariants such as the
Atiyah--Patodi--Singer eta-invariant, the Chern--Simons invariant, or the Adams $e$-invariant.
If the eleven-dimensional manifold itself has a~boundary, the resulting ten-dimensional manifold can be viewed as
a~codimension two corner.
The partition function in this context has been studied by the author in relation to index theory for manifolds with
corners, essentially on the product of two intervals.
In this paper, we focus on the case of framed manifolds (which are automatically Spin) and provide a~formulation of the
ref\/ined partition function using a~tertiary index invariant, namely the $f$-invariant introduced by Laures within
elliptic cohomology.
We describe the context globally, connecting the various spaces and theories around M-theory, and providing a~physical
realization and interpretation of some ingredients appearing in the constructions due to Bunke--Naumann and Bodecker.
The formulation leads to a~natural interpretation of anomalies using corners and uncovers some resulting constraints in
the heterotic corner.
The analysis for type IIA leads to a~physical identif\/ication of various components of eta-forms appearing in the formula
for the phase of the partition function.}


\Classification{81T50; 55N20; 58J26; 58J22; 58J28; 81T30}

\vspace{-2mm}

\section{Introduction}

The goal of this paper is to combine the appearance of corners with that of elliptic cohomology to describe global
aspects of the partition function in M-theory, which we hope could help shed some light on the role of elliptic
cohomology in physics.
Topological study of M-theory is often facilitated by taking it as a~boundary.
Furthermore, the heterotic theory is essentially a~boundary of M-theory.
Then, considering topological aspects in this setting requires the study of a~twelve-dimensional theory whose boundary
theory itself admits a~boundary, i.e.\
forms a~corner of codimension two.
The partition function using index theory for manifolds with corners is analyzed in~\cite{DMW-corner}.
On the other hand, the study of anomalies in M-theory and string theory suggests connections to elliptic cohomology.
In the case of heterotic string theory this has a~long history, in particular in connection to elliptic
genera~\cite{Kil, SW, WittenLNM}.
More recently, by interpreting various anomaly cancellation conditions as orientations with respect to generalized
cohomology theories, direct connections between elliptic cohomology, on the one hand, and M-theory and type~II string
theory, on the other hand, are uncovered~\cite{KS1, KS3, KS2, S-elliptic, tcu, SSSIII}.

It is natural to ask how the above two descriptions can be consistently merged together.
We advocate that the combination of the two pictures, namely that of codimension two corners and that of elliptic
cohomology, f\/its nicely into a~coherent structure form the mathematical point of view.
We implement a~unif\/ied view which we use to study some topological aspects of M-theory in this context for framed
manifolds.
This is done via the $f$-invariant, a~tertiary invariant introduced in~\cite{Lau} and connected to index theory
in~\cite{BN, Bod}.
Some aspects of M-theory on framed manifolds in the context of elliptic cohomology are considered in~\cite{String}.
Here, in addition to extending the relation further to the heterotic theory, we also consider the ef\/fective action and
partition function of type IIA string theory.

{\bf Why framed manifolds.} We would like to encode corners together with cobordism in the context of M-theory.
By now it is established that cobordism invariants corresponding to various structures appear in the construction of the
partition function in M-theory~\cite{DFM, DMW, KSpin, String, DMW-Spinc}.
Structures described previously include Spin, Spin${}^c$ and String structures.
These are related to K-theory and elliptic cohomology.
In this paper we will consider manifolds with a~more basic structure, namely a~framing of the tangent bundle, which
include the class of parallellizable manifolds.
By the Pontrjagin--Thom construction, this amounts to dealing with the sphere spectrum, which in turn can be studied by
means of elliptic cohomology.
We, therefore, provide another angle on the proposals in~\cite{KS1,KS3, KS2, tcu}.
On the other hand, Lie groups provide an interesting class of framed manifolds, and so taking spacetime to be a~Lie
group (or some quotient thereof) resembles~-- and to some extent subsumes~-- Wess--Zumino--Witten models.
Note, however, that being framed automatically means being Spin, so that our discussion certainly includes the
structures that are expected from a~physics point of view, namely Spin structures (see~\cite{DMW-Spinc} for an extensive
description, with an emphasis on the geometry).

{\bf Why framed in the heterotic theory.} We would like to concentrate on the case when the 10-dimensional heterotic
corner of M-theory admits a~framing.
String theory on paralleli\-zable backgrounds is exactly solvable, and hence such backgrounds play a~prominent role in the
theory.
In~\cite{FKY, GPRS, KY, SSJ} a~classif\/ication of (simply-connected) supersymmetric parallelizable backgrounds of
heterotic string theory is given.
For heterotic backgrounds without gauge f\/ields the dilaton is linear and hence can be described by a~Liouville theory,
and the geometry is that of a~parallelized Lie group and hence can be described by a~WZW model.
These include products of Minkowski spaces with the odd-dimensional spheres $S^3$ and $S^7$ and the Lie group SU(3).
For example, for the latter, the ten-dimensional manifold is ${\mathbb R}^{1,1}\times\text{SU}(3)$.
In the presence of nonzero gauge f\/ield strength, the geometry may be deformed away from that of a~group manifold; it is
still parallelizable but with respect to a~metric connection with a~skew-symmetric torsion~\cite{GPR}.
Flux compactif\/ications on group manifolds in heterotic string theory are considered in~\cite{BBDGS}.
These include the group manifolds with zero Euler characteristic underlying ungauged WZW models, such as $S^3 \times
S^1$~\cite{SSTv}.
Examples with nonozero Euler characteristic include connected sums of $\text{SU}(2)\times\text{SU}(2)\cong S^3 \times
S^3$, which admit a~complex structure, but are non-K\"ahler, and have a~nowhere-zero holomorphic form.

{\bf Index-theoretic invariants in various dimensions.} Various index-theoretic invariants appear in the description of
the ef\/fective action, and hence of the partition function, in M-theory.
These arise in the form of an index in the twelve-dimensional extension of M-theory, a~mod~2 index in type IIA in ten
dimensions, and a~secondary invariant in eleven-dimensional M-theory.
Our point of view provides and supports a~dimensional hierarchy of the form
\begin{itemize}\itemsep=0pt
\item
{\it Dimension $4$ and $12$}: The ef\/fective action is given by indices of twisted Dirac operator (see~\cite{flux}).
This is a~main point where topology enters.
Some ref\/inements to elliptic genera appear in~\cite{KS1, String}.

\item
{\it Dimension $3$ and $11$}: The ef\/fective action involves the eta-invariant, the $e$-invariant, or the Chern--Simons
invariant (see~\cite{DMW-Spinc} for an extensive discussion).
In the presence of corners, the Melrose $b$-calculus is used to replace the eta-invariant with the
$b$-eta-invariant~\cite{DMW-corner}.
Some elliptic ref\/inements, in the sense of~\cite{BN-eta, Gal}, are discussed in~\cite{String}.

\item
{\it Dimension~$2$ and $10$}: The ef\/fective action and partition function of type II target string theory and of the
worldsheet theory involve the Arf invariant~\cite{DMW}.
Elliptic ref\/inements of the mod~2 index include the Ochanine invariant as discussed in detail in~\cite{OP2}.
\end{itemize}

Including the case of framed manifolds, we will advocate the structures in the following table

\begin{table}[h!]\centering
\small
\begin{tabular}
{|@{\,\,}c@{\,\,}|@{\,\,}c@{\,\,}|@{\,\,}c@{\,\,}|@{\,\,}l@{\,\,}|@{\,\,}l@{\,\,}|}
\hline
Dimension & Physical Theory & Index invariant & Cohomology Theory & Underlying Structure\tsep{2pt}\bsep{2pt}
\\
\hline
\hline
10 & IIA & $d$ & KO-theory & Closed Spin manifold\tsep{2pt}\bsep{2pt}
\\
\hline
11 & M-theory & $e$ & K-theory & Manifold with boundary\tsep{2pt}\bsep{2pt}
\\
\hline
10 & Heterotic & $f$ & Elliptic cohomology & Manifold with codim-2 corner\tsep{2pt}\bsep{2pt}
\\
\hline
\end{tabular}
\end{table}

Here $d$ is the Adams $d$-invariant, which is the degree (Hurewicz map) for KO-theory and is (a variant of) the mod~2
index of the Dirac operator\footnote{Note that Adams def\/ined the degree~$d$ for any (generalized) cohomology theory.
It is the one based on KO-theory which can be interpreted as an index mod~2.
On the other hand, the one based on integral cohomolo\-gy~$H{\mathbb Z}$ is the `correct' one from the chromatic point of
view (i.e.\
it detects 0th f\/iltration phenomena), but does not carry `nontrivial' information.}, $e$ is the Adams $e$-invariant, and~$f$ is the invariant of Laures for manifolds with corners of codimension two in the context of elliptic cohomology.
The latter is related to the elliptic genus in a~manner similar to how the $e$-invariant is related to the Todd genus in
K-theory.
In the generalization of the $e$-invariant, which takes values in ${\mathbb Q}/{\mathbb Z}$, to modular forms one notes
the following: Since ${\mathbb Q}/{\mathbb Z}$ is not a~ring then it does not make sense to consider ``modular forms
with coef\/f\/icients mod~${\mathbb Z}$''.
Hence, one has to consider modular forms with values in an appropriate ring, which turns out to be the ring of divided
congruences $D$~\cite{Lau}.
More precisely, unlike the 1-line in the Adams--Novikov spectral sequence (ANSS) the 2-line is not cyclic in each
dimension, hence one needs more copies of ${\mathbb Q}/{\mathbb Z}$, and a~good way to do so is via $D \otimes {\mathbb
Q}/{\mathbb Z}$.

{\bf Which elliptic cohomology theory?} Topological modular forms (TMF), while can be viewed as a~sort of a~`universal
elliptic cohomology theory', suf\/fers from a~shortcoming, namely that it is not complex-oriented.
The latter is desirable when dealing with physics (see~\mbox{\cite{KS1, KS3}}).
Therefore, we will consider versions of TMF which are complex-oriented.
A prominent example is TMF$_1(N)$ attached to the universal curve over the ring of integral modular forms for the
congruence subgroup $\Gamma=\Gamma_1(N)$ of ${\rm SL}(2,{\mathbb Z})$.
More precisely, TMF$_1(N)$ is formed of global sections of a~sheaf of spectra over the moduli space of elliptic curves
with level structure; see~\mbox{\cite{Lau, MR}}.
Although elliptic cohomology of level~2 allows to extract the information on the $e$-invariant, its use is by no means
a~requirement as, in fact, KO-theory is enough for that purpose.
The important point is that TMF$_1(N)$ provides natural {\it refinements} of ``well-known'' invariants, such as $d$ and
$e$, as well as entirely new ones, such as~$f$.
An appropriate value for~$N$ turns out to be~3.
In K-theory at times one works away from powers of~2, and the analogy here is working in elliptic cohomology away from
powers of~3.
In fact, if one does not want~2 to be inverted then the smallest level at which this occurs is~3.
Detecting mod~3 phenomena, i.e.\
considering the case $N=3$, connects to anomalies at the prime 3, studied in~\cite{ES2, KSpin}.
The congruence subgroup $\Gamma_1(3)$ appears elsewhere physics, e.g.\
in the context of topological string theory~\cite{AK}.
Note, however, that one should exercise caution in that working at a~f\/ixed prime level is {\it not} equivalent to
focusing on phenomena (e.g.\
anomalies) at that prime.

{\bf Elliptic genera for the heterotic string.} We recall an explicit instance where elliptic genera appear in the
heterotic theory, which we will later view as a~corner.
The modular invariance violating terms can be factored out of the character-valued partition function, which has the
form~\cite{LNSW}
\begin{gather*}
A(q, F, R)= \exp\left[ -\tfrac{64}{\pi^4} G_2(\tau) \big(\operatorname{Tr}F^2 - \operatorname{Tr}R^2\big) \right] \widetilde{A}(q,F,R),
\end{gather*}
where $\widetilde{A}(q,F,R)$ is a~fully holomorphic and modular invariant of weight $-4$, and can be expressed in terms
of the Eisenstein functions $G_4$ and $G_6$ (or equivalently, using $E_4$ and $E_6$), as opposed to the function $G_2$
which is not modular.
From modular invariance, the anomaly always has a~factorized form and is given by the constant term in the elliptic
genus
\begin{gather*}
I_{12}(F,R)=A(q,F,R)\big|_{\text{12-form~coef\/f.~of~}q^0}
= \tfrac{1}{2\pi} \big(\operatorname{Tr}F^2 - \operatorname{Tr}R^2\big) \wedge X_8(F, R),
\end{gather*}
where $X_8(F, R)$ is the Green--Schwarz
polynomial corresponding to the tangent bundle and gauge bundle with curvatures $R$ and $F$, respectively.

{\bf Oultine.} What we do in this paper can be summarized in the following:

1.~We provide a~setting for framed manifolds in M-theory and string theory, starting at the beginning of
Section~\ref{Sec framed}.
We specialize to parallelizable manifolds, stably parallelizable manifolds, and in particular to Lie groups and
homogeneous spaces in Sections~\ref{Sec par},~\ref{Sec st}, and~\ref{Sec Lie}, respectively.
Since M-theory involves boundaries, we describe how framed manifolds arise as boundaries in our context in
Section~\ref{Sec bound}.

2.~We describe framed cobordism invariants in connection to M-theory and string theory.
After a~general discussion on framed cobordism in Section~\ref{Sec fra}, we describe the relation between 10-dimensional
string theory and the $d$-invariant (which is a~variant of the Arf invariant) in Section~\ref{Sec d}, and then the
relation between the $e$-invariant and 11-dimensional M-theory in Section~\ref{Sec e}.
Along the way we explain the ef\/fect on the partition function, and in Section~\ref{Sec ch} we consider that from the
point of view of change of framing.
We also describe the parity symmetry of the $C$-f\/ield in this context.

3.~Having described both framed structures and invariants, we introduce the corner formulation and show in
Section~\ref{Sec build} how the various interconnected theories, together with their boundaries and some dualities, f\/it
nicely into the structure of framing and corners.
This provides a~transparent view on how K-theory and elliptic cohomology enter into the setting.
For instance, the formulation of the $C$-f\/ield in M-theory in~\cite{DFM}
can be cast in this setting in a~very natural way.

4.~We consider type IIA string theory on a~manifold with boundary in connection to M-theory, itself with a~boundary and
as a~boundary.
Since the phase of the partition function is given by an index, it is natural to describe the partition function, and
especially the phase, in this index-theoretic context.
This allows us to get an expression of the phase and provide an interpretation of eta-forms appearing in~\cite{MaS},
thereby extending similar interpretations in~\mbox{\cite{S-gerbe, DMW-corner}}.
This also serves as a~warm up via secondary index theory for the application of the tertiary index theory in later sections.

5.~We consider the formulation of the heterotic theory as a~corner in Section~\ref{Sec as}.
The factorization of the anomaly can be viewed via the splitting of the tangent bundle in the context of framing.
Furthermore, we show in Section~\ref{Sec an} that the general form of the anomaly as well as the cancellation of that
anomaly point to the presence of corners.
This can be generalized to anomaly cancellation as a~general process.
Then, in Section~\ref{Sec 1}, we lift the one-loop term to twelve dimensions and then consider the reduction to the
corner.
This results in a~constraint on degree twelve Chern numbers that generalize the constraint on degree ten Chern numbers
in~\cite{DMW}, and gives rise to a~cup product composite Chern--Simons theory.
We illustrate how these conditions af\/fect the corner by highlighting the example of the ten-dimensional Lie group Sp(2)
in~\cite{Lau} which turns out to be the physically important one (see~\cite{ES2}).

6.~We next combine the framing and corners description of the heterotic string with the elliptic cohomology aspect via
the framed cobordism invariant at chromatic level~2, namely the $f$-invariant of Laures~\cite{Lau}, and its geometric
ref\/inement by Bunke--Naumann~\cite{BN} and Bodecker~\cite{Bod}.
This can be viewed as the reduction of the index~-- i.e.\
the topological part of the action~-- from twelve dimensions to the corner.
In Section~\ref{Sec tmf} we highlight the connection to topological modular forms and Tate K-theory and how the terms in
the ef\/fective action get ref\/ined to $q$-expansions, as in~\cite{String}.
As we explain in Section~\ref{Sec f}, the $f$-invariant captures the nonzero $q$ part of the expansion and hence, in the
view of treating $q$ as a~sort of a~`coupling constant'~\cite{S-elliptic}, the quantum aspects of the theory.
We highlight this in the example of $S^3 \times S^7$.

We emphasize exposition to explain the various interconnections between the mathematical constructions and the physical
ingredients and settings.

\section{Framed manifolds and framed bundles in M-theory}
\label{Sec framed}

In this section we describe relevant classes of framed manifolds which appear in our setting.
These include parallelizable manifolds such as Lie groups and certain homogeneous spaces.
We then highlight the relevance of structures on these spaces to the physics in M-theory and string theory.
We will denote by ``admissible manifolds" those manifolds that can be taken as spaces on which string theory or M-theory
can be compactif\/ied, with or without f\/luxes.

{\bf Framed manifolds.} If $M$ is a~closed $n$-dimensional manifold with tangent bundle $TM$, then its stable tangent
bundle\footnote{Note that this def\/inition involves (commonly accepted) abuse of notation.
Strictly speaking, the stable tangent bundle of $M$ should be thought of as an equivalence class of such, namely as
a~class in reduced real K-theory $\widetilde{\text{KO}}(M)$.
A similar remark holds for the stable framing.} $T^{\text{st}}M$ is the direct sum of $TM$ with a~large\footnote{That is,
large $r$.
However, the exact value of $r$ will not be important for us.} trivial bundle $M \times {\mathbb R}^r$
\begin{gather*}
T^{\text{st}}M \cong TM \oplus (M \times {\mathbb R}^r).
\end{gather*}
A (stable) framing on $M$ means a~trivialization $f$ of $T^{\text{st}}M$, that is, a~set $f=(f_1, \dots, f_{n+r})$ of
$(n+r)$-sections of $T^\text{st}M$, linearly independent everywhere.
One can consider this from the point of view of embeddings.
A framing on a~manifold~$M^n$ smoothly embedded in Euclidean space~${\mathbb R}^{n+k}$ consists of an ordered set of
vectors $\{v_1(x), \dots, v_k(x)\}$ varying smoothly with \mbox{$x\in M^n$} and providing a~basis for the normal space of $M^n$
in~${\mathbb R}^{n+k}$ at~$x$.
In terms of classifying spaces of $G$-structures (see~\cite{Sto}), a~framing on a~smooth manifold $M$ is a~pair $(h,
\tilde{\nu})$ such that $h: M\to {\mathbb R}^{n+k}$ is an embedding with normal bundle classif\/ied by a~map $\nu: M \to
\text{BO}(k)$ with a~lifting $\tilde{\nu}: M \to \text{EO}(k)$, where $\text{EO}(k)$ is the total space of the universal principal $\text{O}(k)$
bundle
\begin{gather*}
\xymatrix{ & \text{EO}(k) \ar[d]^{\text{pr}_k}
\\
M \ar[ur]^{\tilde{\nu}} \ar[r]^{\nu} & \text{BO}(k) }
\end{gather*}
with $\text{pr}_k: \text{EO}(k) \to \text{BO}(k)$ being the bundle projection.

{\bf Framed vector bundles.} A rank $r$ vector bundle $E \to M$ is called trivial or trivializable if there exists
a~bundle isomorphism $E \cong M \times {{\mathbb R}}^r$ with the trivial rank $r$ bundle over $M$.
A bundle isomorphism $E \to M \times {{\mathbb R}}^r$ is called a~trivialization of $E$, while an isomorphism $\varphi:
M \times {{\mathbb R}}^r \to E$ is called a~framing of $E$.
Denote by $(e_1, \dots, e_r)$ the canonical basis of the vector space ${\mathbb R}^r$, and regard the vectors $e_i$ as
constant maps $M \to {\mathbb R}^r$, i.e.~as sections of $M \times {{\mathbb R}}^r$.
The isomor\-phism~$\varphi$ determines sections $f_i=\varphi (e_i)$ of $E$ with the property that for every $x\in M$ the
collection $(f_1(x), \dots, f_r(x))$ is a~{\it frame} of the f\/iber $E_x$.
This shows that we can regard any framing of a~bundle $E \to M$ of rank $r$ as a~collection of $r$ sections $\{u_1,
\dots, u_r\}$ which are pointwise linearly independent.
Thus one has that a~pair ``(trivial bundle, trivialization)" deserves to be called a~{\it trivialized}, or {\it framed}
bundle.

In the following we provide what might be viewed as a~toolkit for admissible manifolds, whereby we provide an extensive
class of examples.

\subsection{Parallelizable manifolds}
\label{Sec par}
A parallelizable manifold is a~manifold whose tangent bundle is trivial.
A trivialization provides a~framing in a~natural way.
It is important to emphasize that, in general, a~trivial bundle is not canonically trivialized, a~fact which is the
source of anomalies from the geometric point of view.

As every parallelizable manifold is Spin, it is an admissible manifold (in a~strong sense) in M-theory.
Such manifolds are often decomposable.
The product of two parallelizable mani\-folds is also parallelizable.
However, the product of two stably parallelizable mani\-folds is not necessarily parallelizable.
However, a~product $M \times N$ is parallelizable if and only if $M$ and $N$ are stably pa\-ral\-lelizable and either factor
has a~vanishing Euler characteristic, since \mbox{$\chi (M \times N)=\chi (M)\cdot \chi (N)$}.
This is automatically satisf\/ied if the total dimension is odd.

{\bf Examples.} {\bf 1.~Lie groups.} All Lie groups (and their quotients by f\/inite subgroups) are parallelizable.
We discuss this important class of examples in detail in Section~\ref{Sec Lie}.

{\bf 2.~(Projective) Stiefel manifolds.} The real and complex Stiefel manifolds $V_{n,k}$ are pa\-ral\-lelizable if $k\geq 2$.
For complex Stiefel manifolds one has the following (see~\cite{AGMP}).
$PV_{n,k}$ is the quotient space of the free circle action on the complex Stiefel manifold $V_{n,k}$ of orthonormal
$k$-frames in complex $n$-space given by $z(v_1, \dots, v_k)=(zv_1, \dots, zv_k)$.
If $k<n-1$ then $PV_{n,k}$ is not stably parallelizable.
The manifold $PV_{n,n-1}$ is parallelizable, except $PV_{2,1}=S^2$, while $PV_{n,n}$ is the projective unitary group,
and so is parallelizable.

{\bf 3.~Grassmannian manifolds.} The only real Grassmannian manifolds $\text{Gr}_k({\mathbb R}^n)$ which are parallelizable are
the obvious cases: $\text{Gr}_1({\mathbb R}^2)\cong {\mathbb R} P^1$, $\text{Gr}_1({\mathbb R}^4)\cong \text{Gr}_3({\mathbb
R}^4)\cong {\mathbb R} P^3$ and $\text{Gr}_1({\mathbb R}^8)\cong \text{Gr}_7({\mathbb R}^8) \cong {\mathbb R} P^7$.
The ten-dimensional manifold $X^{10} = {\mathbb R} P^3 \times {\mathbb R} P^7$ plays an important role in the subtle
aspects of K-theoretic description of the f\/ields in type II string theory~\cite{BEJMS}.

{\bf 4.~Homogeneous spaces.} Many homogeneous spaces are known to be parallelizable: Lie groups, Stiefel manifolds, quotients of
the form $G/T$ where $G$ is a~Lie group and $T$ is a~non-maximal toral subgroup.
This provides many examples involving the relevant low-rank Lie groups.
Another relevant class of homogeneous spaces is the following.
Let $G=\text{SU}(n)$ and $H=\text{SU}(k_1) \times \dots \times \text{SU}(k_r)$, $r=\text{rank}(G)$, embedded in $G$ in an
arbitrary manner.
For an appropriate choice of $\{k_i\}$, if $G/H$ is parallelizable then $G/H$ is either a~complex Stiefel manifold or is
of the form $\text{SU}(n)/\text{SU}(2)^{\times k}$ where the subgroup $\text{SU}(2)^{\times k}$, $2k \leq n$ is embedded in
the standard fashion~\cite{Si}.
Most relevant for us is the nine-dimensional manifold $\text{SU}(4)/(\text{SU}(2) \times \text{SU}(2))$.

{\bf 5.~Products of spheres.} The product of a~sphere with a~sphere of odd dimensions is always parallelizable (see~\cite{Ker}).
For example, let us consider the eleven-dimensional manifold $Y^{11}=S^4 \times S^7$, which is important in the f\/lux
compactif\/ication of M-theory.
The 7-sphere $S^7$ admits a~nowhere zero section, so that the tangent bundle is the sum $TS^7=\eta \oplus \varepsilon^1$
for some rank 6 bundle $\eta$.
Let $\text{pr}_1$ and $\text{pr}_2$ denote the projections of the product to the f\/irst and second factors, respectively.
Then $T(S^4 \times S^7)=\text{pr}_1^*(TS^4) \oplus \text{pr}_2^*(\eta \oplus \varepsilon^1)$; now the second summand gives
$\text{pr}_2^*(\eta)\oplus \varepsilon^1$ and so using $\text{pr}_1^*(TS^4)\oplus \varepsilon^1=\text{pr}_1^*(TS^4 \oplus
\varepsilon^1)=\varepsilon^5$, in total we have $\text{pr}_2^*(\eta \oplus \varepsilon^5)=\varepsilon^{11}$, which shows
that indeed $S^4 \times S^7$ is parallelizable.
Similar remarks hold for the decomposable eleven-dimensional manifolds $S^8 \times S^3$, $S^6 \times S^5$, $S^2 \times
S^9$, and $S^1 \times S^{10}$.

{\bf 6.~Principal bundles over parallelizable manifolds.} If the base space of a~principal bundle is parallelizable then so is
the total space, since the f\/iber is isomorphic to a~Lie group, which is parallelizable.
In relating M-theory in eleven dimensions to various favors of string theory in lower dimensions, one performs
dimensional reduction, which can usually be viewed as a~reduction of the total space of a~principal bundle to its base.

\subsection{Stably parallelizable manifolds}
\label{Sec st}

A~bundle $E$ is said to be {\it stably trivial} if its Whitney sum with a~trivial bundle is trivial.
A~manifold $M$ is said to be {\it stably parallelizable} or a~{\it $\pi$-manifold} if the tangent bundle $TM$ is stably
trivial.
Note that if $TM\oplus \varepsilon^k$ is trivial then $TM \oplus \varepsilon^1$ is already trivial.
If a~connected, stably parallelizable manifold~$M$ has {\it non-empty} boundary, then it is actually
parallelizable~\cite{KM}.

{\bf Reduction of structure group for stably parallelizable manifolds.} Let $Y^{11}$ be a~connected stably
parallelizable closed eleven-dimensional manifold.
There is, up to isomorphism, exactly one stably trivial, but not trivial, 11-dimensional vector bundle $\tau$ over
$Y^{11}$.
It may be described as the pullback of the tangent bundle of $S^{11}$ by a~map $f: Y^{11} \to S^{11}$ of degree one
(collapsing the complement of an open disk)~\cite{CG}.
It follows from~\cite{Th} that the structure group of $\tau$ can be reduced to $\text{SO}(k)$ by the standard inclusion $\text{SO}(k)
\hookrightarrow \text{SO}(11)$ if and only if $12 \equiv 0$ mod $a(12-k)$, where $a(r)$ is the Hurwitz--Radon number of $r$.
The special case in which $\tau$ is the tangent bundle  is considered in~\cite{BK}.

{\bf Properties.} (Stably) parallelizable manifolds enjoy the following useful properties.

{\bf 1.}~The boundary of a~parallelizable manifold is a~$\pi$-manifold.
This will be useful when considering various boundaries and corners.

{\bf 2.}~The product of two $\pi$-manifolds is a~$\pi$-manifold.
This will be useful when we consider decomposable $\pi$-manifolds, which will be the main class of admissible manifolds.

{\bf 3.}~Every stably parallelizable manifold is Spin.
This also follows from the more general fact that framed manifolds are Spin.
Hence such manifolds are physically admissible.

{\bf 4.}~Suppose $H
\subset K \subset G$ is a~sequence of closed Lie groups.
If $G/H$ is stably parallelizable then so is $K/H$.

{\bf Examples.} {\bf 1.~All parallelizable manifolds.} This includes spheres in dimensions 1, 3, and 7.
On the other hand, the only real Grassmannian manifolds $G_k({\mathbb R}^n)$ which are stably parallelizable are the
parallelizable ones, as in Section~\ref{Sec par} above.

{\bf 2.~Spheres.} While spheres are stably parallelizable.
They have many interesting properties including $TS^n \oplus \varepsilon^1 = \varepsilon^{n+1}$.

{\bf 3.~Homogeneous spaces.} An example which is not (strictly) parallelizable is the following.
Let $G$ be a~simple 1-connected compact Lie group and $H$ a~closed connected subgroup.
Then $G/H$ is stably parallelizable if and only if the adjoint representation $\text{Ad}_H$ of $H$ is contained in the
image of the restriction map of real representation rings $\text{RO}(G) \to \text{RO}(H)$~\cite{SiW}.
A homogeneous space which is almost parallelizable not strictly parallelizable is $G/T_\text{max}$, the quotient of a~Lie
group $G$ by a~maximal torus.

{\bf 4.~Sphere bundles over $\boldsymbol{\pi}$-manifolds.} Recall that the tangent bund\-le of a~sphere bundle $S(E)$ takes the form
$TS(E)\cong \pi^* TM^n \oplus T_F(S(E))$, with the canonical isomorphism $1\oplus T_F(S(E))\cong \pi^*E$, where $T_F$ is
the vertical tangent bundle.
Assume that $M^n$ is a~$\pi$-manifold.
Then $S(E)$ is a~$\pi$-manifold if $\pi^*E \to S(E)$ is stably trivial.
In particular, $S(E)$ is a~$\pi$-manifold if $M^n$ is the $n$-sphere; see~\cite{Sm}.

{\bf 5.~Sphere bundles over spheres.} This is a~class of examples that will be very useful for us when considering the partition
function in later sections.
Let $E \to M^n$ be a~smooth oriented $m$-plane bundle with associated sphere bundle $ S^{m-1} \to S(E) \to M^n $ in some
Riemannian metric.
We specify to $M^n=S^n$ and introduce the disk bundle $\mathbb{D}^m \to \mathbb{D}(E) \to S^n$ associated to a~vector
bundle $E$ over the sphere $S^n$.
Recall that $T(\mathbb{D}(E))\cong \pi^*TS^n \oplus \pi^*E$, so that if $\phi$ is a~stable trivialization of $E$ there
is induced a~stable framing of $\mathbb{D}(E)$ by pulling back $\phi$ and the usual stable framing of $TS^n$ along
$\pi$.
Note that $\partial (\mathbb{D}(E); \phi)=(S(E); \phi)$ where $\phi$ is the stable framing of $S(E)$.
In general, such bundles are only stably parallelizable.

{\bf Extending almost parallelizable to parallelizable manifolds.} Here we will contrast the case of string theory and
M-theory, in the sense of even- vs.
odd-dimensional manifolds.
If the dimension of~$M$ is even, the parallelizability of a~stably parallelizable manifold is determined by the
vanishing of the Euler characteristic of~$M$.
Thus if the Euler characteristic is zero then any stably parallelizable manifold is in fact parallelizable.
For us this includes ten-dimensional manifolds appearing in type IIA and heterotic string theory.
On the other hand, if the dimension of~$M$ is odd, $M$ is parallelizable if and only if its Kervaire semi-characteristic
$\chi_{\frac{1}{2}}(M)$, def\/ined via mod~2 homology by
\begin{gather}
\chi_{\tfrac{1}{2}}(M)= \tfrac{1}{2}
\sum\limits_{i=0}
^{\lfloor \dim (M)/2 \rfloor} \dim_{{\mathbb Z}_2} H_i(M;{\mathbb Z}_2)
\qquad
({\rm mod}\; 2)
\label{semi}
\end{gather}
vanishes.
Therefore, similarly, when $\chi_{\frac{1}{2}}(M)=0$ then a~stably parallelizable manifold becomes parallelizable.
This places a~condition on the homology of the manifolds; see Section~\ref{Sec RS}.

\subsection{Lie groups and homogeneous spaces as framed manifolds}
\label{Sec Lie}

Lie groups form an interesting class of examples of compactif\/ication manifolds which are able to carry subtle torsion
information about f\/ields in spacetime.
See e.g.~\cite{MMS} for a~description of such WZW models in the context of twisted K-theory.
The discussion we give below, together with the construction of twisted Morava K-theory in~\cite{SaW}, allows for an
extension to detect f\/iner invariants (see Section~\ref{Sec fra}).

{\bf Framings on Lie groups.} The left invariant vector f\/ields of a~compact Lie group $G$ induce a~specif\/ic isomorphism
$\ensuremath{\mathcal L}$, the left invariant framing, between the tangent bundle of $G$ and the product bundle $G
\times {\mathbb R}^{\dim G}$.
Indeed, the tangent bundle $T(G)$ of any Lie group $G$ is trivial.
Take a~basis $\{ e_1, \dots, e_n\}$ of the tangent space at the origin $T_e(G)$, where $n=\dim G$.
Denote by $R_g$ the right translation by $g$ in the group def\/ined by $ R_g: x \mapsto x\cdot g$, for all $x \in G$.
This is a~dif\/feomorphism with inverse $R_{g}^{-1}=R_{g^{-1}}$ so that the dif\/ferential $DR_g$ def\/ines a~linear
isomorphism $DR_g: T_e(G) \to T_gG$.
Since the multiplication $G \times G \to G$, given by $(g,h) \mapsto g\cdot h$, is a~smooth map then the vectors $
f_i(g)= DR_g(e_i) \in T_gG$, $i=1, \dots, n$, def\/ine smooth vector f\/ields over $G$.
Then for every $g\in G$ the set $\{ f_1(g), \dots, f_n(g)\}$ is a~basis of $T_g(G)$ so that there is indeed a~vector
bundle isomorphism $ \phi: G \times {\mathbb R}^n \to TG$ taking $(g; e^1, \dots, e^n)$ to $(g;
\sum e^i f_i(g))$.
Similarly, the same holds for framing via the left translation $\ensuremath{\mathcal L}_g$.
Then the right invariant framing $ \ensuremath{\mathcal R}: T(G)\cong G \times T_e(G) $ of $G$ is given by
$\ensuremath{\mathcal R}(v)=(g, R_{g^{-1}
}(v))$ where $v \in T_g(G)$.

{\bf Framings on homogeneous spaces $\boldsymbol{G/H}$.} Let $G$ be a~compact connected Lie group and~$H$ a~closed subgroup of~$G$.
Let $T(G/H)$ denote the tangent bundle bundle of the coset~$G/H$.
Consider the $H$-principal bundle $H \to G \buildrel{\pi}\over{\to} G/H$.
Then the tangent bundle of $G$ decomposes as $T(G) \cong \pi^*T(G/H) \oplus T_H(G)$.
This isomorphism is compatible with the right $H$-action, and so there is an isomorphism of vector bundles over $G/H$,
namely $T(G)/H \cong T(G/H) \oplus T_H(G)/H$.
Let $\text{ad}_H$ denote the adjoint representation of $H$ on $T_e(H)$.
Similarly, there is an isomorphism $T_H(G)/H\cong G\times_H T_e(H)$ of vector bundles over~$G/H$, where~$H$ acts
on~$T_e(H)$ via~$\text{ad}_H$.
Combining the above bundles gives the isomorphism of vector bundles over~$G/H$
\begin{gather}
G/H \times T_e(G)\cong T(G/H) \oplus G \times_{\text{ad}_H}T_e(H).
\label{gh}
\end{gather}
So if $\text{ad}_H$ is contained in the image of the restriction map $\text{RO}(G)\to \text{RO}(H)$ of real representation rings
then~\eqref{gh} gives a~framing of $G/H$ (see~\cite{LS}).

\subsection{Framed boundaries}
\label{Sec bound}

We will consider framed manifolds in twelve, eleven, ten, and nine dimensions, and in the last three cases we would like
to allow the manifolds to be boundaries.
We will consider restrictions for this to occur and illustrate with useful examples.

{\bf Lie groups.} Let $G$ be a~compact Lie group.
Two natural questions that arise in our context are: When is $G$ the boundary of a~compact manifold $Z$? In this case,
when is $Z$ parallelizable? We highlight two cases that are important to our discussion:
\begin{enumerate}\itemsep=0pt
\item
Disk bundles ${\mathbb D}(\ensuremath{\mathcal L}_{\mathbb C})$ of the canonical complex line bundles
$\ensuremath{\mathcal L}_{\mathbb C}$ over the quotient $G/S^1$.
The boundary $Y$ of the total space $Z$ of the complex line bundle is the circle bundle $S(\ensuremath{\mathcal L})$
given by $G \to G/S^1$ with $\partial Z=G$.
This context (for manifolds that are not necessarily framed) is discussed extensively in~\cite{DMW-Spinc}.

\item
Similarly for disk bundles ${\mathbb D}(\ensuremath{\mathcal L}_{\mathbb H})$ of the quaternionic line bundles
$\ensuremath{\mathcal L}_{\mathbb H}$ over $G/\text{SU}(2)\cong G/S^3$.
The boundary $Y$ of the total space $Z$ of the quaternionic line bundle is the sphere bundle $S(\ensuremath{\mathcal
L}_{\mathbb H})$ given by $G \to G/S^3$ with $\partial Z=G$.
The example we have in mind in this case is the group ten-manifold $\text{Sp}(2) \cong \text{Spin}(5)$ or SO(5) and their
quotients with f\/inite groups.
See also~\cite{ES2} for an application to D-brane anomalies.
\end{enumerate}

{\bf Generalized f\/lag manifolds.} Let $G$ be a~compact Lie group of rank $l$ and $T$ a~maximal torus.
Then the f\/lag manifold $G/T$ is a~$\pi$-manifold~\cite{BH3}, and an explicit bounding mani\-fold~$W$ with a~corresponding
stable framing can be constructed as follows~\cite{PS}.
Let $\frak{g}$ and $\frak{t}$ be the Lie algebras of $G$ and $T$, respectively.
Under the adjoint action of $T$, $\frak{g}$ decomposes as an $\text{ad}(T)$-module as $\frak{g}=\frak{t}\oplus_\alpha
\frak{g}_\alpha$, where $\alpha$ are certain linear forms $\alpha: \frak{t}\to {\mathbb R}$ and the $\frak{g}_\alpha$
are two-dimensional $T$-modules corresponding to $e^\alpha: T \to \text{SO}(2)$.
The subspace $\frak{C}_\alpha=\frak{t}\oplus \frak{g}_\alpha$ is in fact a~Lie subalgebra of $\frak{g}$ isomorphic to
${\mathbb R}^{l-1}\oplus \frak{su}(2)$, where ${\mathbb R}^{l-1}=\ker \alpha$ is the center and $\frak{su}(2)$ is the
commutator subalgebra of $\frak{C}_\alpha$.
Denote by $C_\alpha \leq G$ the closed connected subgroup corresponding to $\frak{C}_\alpha \leq \frak{g}$.
Then the 2-sphere bundle
\begin{gather}
C_\alpha/T \to G/T \to G/C_\alpha
\label{2-sph}
\end{gather}
has a~corresponding disk bundle $W$ which is stably parallelizable.
The tangent bundle of $W$ is given by $ TW \cong \pi^*T(G/C_\alpha)\oplus \pi^*\xi$, where $\xi$ is the 3-plane vector
bundle corresponding to the 2-sphere bundle~\eqref{2-sph}.
For example, when $G$ is the Lie group $G_2$ the total space of the 2-sphere bundle is a~12-dimensional manifold.

{\bf $\boldsymbol{S^1}$- and $\boldsymbol{S^3}$-action and framing of the disk bundle.}  Consider the case when~$H$ is~$S^1$ or~$S^3$ and $Z$ the
corresponding disk bundle over $G/H$ with projection $p$.
Denote by $\text{Ad}_G$ (resp.~$\text{Ad}_H$) the adjoint representation of $G$ (resp.~$H$).
Then the restriction of $\text{Ad}_G$ to $H$ decomposes~as
\begin{gather}
\text{Ad}_G\big|_H=\text{Ad}_{(G,H)}\oplus \text{Ad}_H,
\label{ad}
\end{gather}
since $\text{Ad}_G|_H$ contains $\text{Ad}_H$ as a~sub-representation.
Let $H$ act via $\text{Ad}_G|_H$ on the tangent space $T_e(G)$, decomposing it via~\eqref{ad}, as $T_e(G)=V \oplus
T_e(H)$.
From $T(G)/H\cong G \times_H T_e(G)$ and $T_H(G)/H\cong G \times_H T_e(H)$, one gets $ T(G/H) \cong G \times_H V $.
Suppose that there is a~real representation $f$ of $G$ such that $f|_H=\text{Ad}_{(G,H)}\oplus
\sigma \oplus \ell$, where the integer $\ell$ denotes the $\ell$-dimensional trivial representation, and $\sigma$ is the
inclusion $H \hookrightarrow \text{SO}(r+1)$ for $r=1,3$.
Applying~$f$ to~$TZ\cong p^*(T(G/H) \oplus \eta$, where $\eta$ is the vector bundle associated via $\sigma$ to the
principal $H$-bundle $\pi: G \to G/H$, yields an isomorphism $ \phi: TZ \oplus (Z\times {\mathbb R}^\ell) \to Z \times
{\mathbb R}^{d+\ell +1}
$, which provides a~framing for~$Z$.
So the framed manifold $(Z, \phi)$ bounds the framed manifold $(G, -f)$.
See~\cite{Mi2} for more details.
More examples can be found in~\cite{AGP}.

We now present an example which is central to our discussion.

{\bf Example. Sp(2).} The ten-dimensional Lie group Sp(2) can be viewed as a~3-sphere bundle over the 7-sphere, $S^3 \hookrightarrow
\text{Sp}(2) \to S^7$.
This example is used in~\cite{ES2} to study D-brane anomalies at the prime $p=3$.
Since this is a~sphere bundle, it is a~boundary of a~disk bundle $\mathbb{D}^4 \hookrightarrow Y^{11}=D(\text{Sp}(2)) \to
S^7$, which is a~Spin manifold.
The framing can also be extended to the disk bundle by the above results, even though $[\text{Sp}(2), \alpha,
\ensuremath{\mathcal L}]=0$ in the stable homotopy group $\pi_{10}^sS^0$ (see Section~\ref{Sec fra}).

{\bf Topological conditions on stably parallelizable manifolds with boundary.} We consider M-theory on an
eleven-manifold $Y^{11}$.
Then the semi-characteristic of $Y$ is def\/ined in~\eqref{semi}.
Let $Z^{12}$ be a~compact $12$-dimensional manifold with boundary $\partial Z^{12}=Y^{11}$.
Then, from the general result in~\cite{BSS}, the Euler characteristic of $Z^{12}$ and the Euler semi-characteristic of
$Y^{11}$ are related as
\begin{gather*}
\chi \big(Z^{12}\big)= \chi_{\frac{1}{2}}\big(Y^{11}\big) \mod 2.
\end{gather*}

\noindent This places a~condition on the cohomology of $Y^{11}$ and of its bounding manifold $Z^{12}$.

\subsection{Framed cobordism}
\label{Sec fra}

We are considering manifolds which can be boundaries and which at the same time admit a~framing.
The natural context to study these is framed cobordism.

\looseness=1
{\bf Framed cobordism classes and the parity symmetry in M-theory.} Let $M_1$ and $M_2$ be two closed $n$-dimensional
framed manifolds.
We say that $M_1$ and $M_2$ are framed cobordant (written $M_1
\sim M_2$) if there are $(n+1)$-dimensional compact framed manifolds $W_1$, $W_2$ with framed dif\/feomorphism $M_1
\coprod \partial W_1 \cong M_2 \coprod \partial W_2$, where $\partial W_1$ and $\partial W_2$ have the induced fra\-mings.
The empty set can be viewed as an $n$-dimensional smooth framed manifold with a~unique fra\-ming.
A framed manifold $(M, f)$ is {\it null-cobordant}
or cobordant to zero if $M$ is the boundary of a~compact manifold $X^{n+1}$ endowed with a~framing $\tilde{f}$ that
restricts on $M$ to $\hat{n} \oplus f$, where $\hat{n}$ is the unit outward-pointing normal f\/ield of $M$ in $X$.
The {\it inverse} $-(M, f)$ of a~framed mani\-fold~$(M, f)$ is def\/ined by taking $M$ with the ``opposite" framing, i.e.\
with the framing obtained by reversing one of the sections of $f$.
Note that this implements the discrete parity symmetry of M-theory on manifolds with vanishing f\/irst Spin characteristic
class, i.e.\
on String manifolds and hence framed manifolds.
This symmetry is given by an odd number (in this case one) of space and time ref\/lections together with a~ref\/lection of
the $C$-f\/ield $C_3 \mapsto - C_3$.
Two framed $n$-manifolds~$(M_1, f_1)$, $(M_2, f_2)$ are framed cobordant if their disjoint union $(M_1, f_1) \cap -(M_2,
f_2)$ is null-cobordant.
This is an equivalence relation for framed manifolds, and the set of equivalence classes of framed $n$-manifolds forms
an abelian group $\Omega_n^\text{fr}$ under disjoint union of manifolds.

\looseness=1
{\bf Framed cobordism in 9, 10, and 11 dimensions.} A closed framed 10-manifold $M^{10}$ represents a~class $[M^{10}]
\in \Omega_{10}^\text{fr}\cong \pi_{10}^s(S^0) \cong {\mathbb Z}_2 \oplus {\mathbb Z}_3 \cong {\mathbb Z}_6$ via the
Pontrjagin--Thom construction.
We will also be interested in the 9-dimensional and 11-dimensional cases, for which $\pi_9^s\cong {\mathbb Z}_2 \oplus
{\mathbb Z}_2 \oplus {\mathbb Z}_2$  and $\pi_{11}^s \cong {\mathbb Z}_7 \oplus {\mathbb Z}_8 \oplus {\mathbb Z}_9$,
respectively.
The fact that these groups are nonzero implies that there are obstructions to having a~framed 9-manifold, 10-manifold,
or 11-manifold to be a~boundary.
However, as in~\cite{DFM} for the Spin case, we will assume that the manifolds that we have are such that there are no
such obstructions, i.e., the boundaries are given to us from the start; we will take situations where we have a~specif\/ic
given boundary.
By their very construction, non-triviality of the framed bordism groups imply that there are obstructions to having
framed manifolds occur as {\it framed} boundaries.
Even when they do not bound framed manifolds, it might be suf\/f\/icient that they bound at least a~physically admissible
manifold, e.g.\
a~Spin manifold.

{\bf Lie groups as elements in framed cobordism.} If $G$ is a~$k$-dimensional compact oriented Lie group then every
trivialization of the tangent bundle gives rise to a~trivialization of the stable normal bundle and hence to an element
of the $k$th framed cobordism group $\Omega_k^\text{fr}$.
If two choices of linear isomorphisms of the Lie algebra $\frak{g}$ with ${\mathbb R}^k$ dif\/fer by an element of ${\rm
GL}(k,{\mathbb R})$ of positive determinant then the corresponding tangential trivializations are homotopic through
trivializations and hence determine the same element of $\Omega_k^\text{fr}$.
Therefore, a~compact oriented $k$-dimensional Lie group gives rise to a~well-def\/ined element $[G]\in \Omega_k^\text{fr}$.

{\bf Adams f\/iltration.} A compact Lie group $G$ with its left invariant framing $\ensuremath{\mathcal L}$ def\/ines, via
the Pontrjagin--Thom construction, an element $[G, \ensuremath{\mathcal L}]$ in the stable homotopy groups of spheres~$\pi_*^s$.
The f\/iltration is a~good measure of the complexity of~$\pi_*^s$.
For a~compact Lie group of rank $r$ the f\/iltration is {\it at least}~$r$.
A result of~\cite{Kn} states that for $G$ a~compact Lie group of rank $r$, the element~$[G, \ensuremath{\mathcal L}]$ in
$\pi_*^s(S^0)$ def\/ined by~$G$ in f\/iltration $r$.
This f\/iltration is the same as the chromatic level.
To detect chromatic phenomena at level 1, that is via K-theory, Lie groups of rank 1 should be used.
However, if we want to detect chromatic phenomena at level~2, corresponding to elliptic cohomology (or to Morava
K-theory K(2)), we should consider Lie groups of rank~2.
Therefore, a~priori, the most relevant groups for us will be e.g.\
$G_2$, Sp(2), SO(5), Spin(5), SU(3), and their quotients.
The element resulting from the Pontrjagin--Thom construction depends only on the orientation of the basis and is denoted
by $[G, \alpha, \ensuremath{\mathcal L}]$, where $\alpha$ is the orientation of $G$.
Using right translation instead leads to the element $[G, \alpha, \ensuremath{\mathcal R}]$.
Sometimes we will leave the orientation out of the notation.

{\bf Examples.} {\bf 1.~Spheres.} We consider the Lie groups which are spheres, namely $S^1\cong \text{SO}(2)$ and $S^3\cong\text{Spin}(3)$.
The elements $[S^1, \ensuremath{\mathcal L}]$ and $[S^3, \ensuremath{\mathcal L}]$ represent the Hopf maps $\eta \in
\pi_1^S={\mathbb Z}_2$ and $\nu \in \pi_3^S={\mathbb Z}_{24}$, respectively.

 {\bf 2.~Tori.} The three-dimensional torus $T^3=S^1 \times S^1 \times S^1$ represents an element $\eta^3\in \pi_3^s$, where
$\eta \in \pi_1^s$ is the generator represented by $S^1$.

  {\bf 3.~Central extensions.} Let $C$ be a~f\/inite central subgroup of $G$ so that there is an extension $C \to G \to G/C$.
It is natural to ask how the classes $[G, \ensuremath{\mathcal L}]$ and $[G/C, \ensuremath{\mathcal L}]$ might be
related.
For example, SO(3) represents $2\nu$ so that $[\text{SO}(3), \ensuremath{\mathcal L}]=2[\text{Spin}(3), \ensuremath{\mathcal
L}]$.
In general $[\text{SO}(2n), \ensuremath{\mathcal L}]=2[\text{Spin}(2n), \ensuremath{\mathcal L}]$ which is zero for $n
\geq 2$~\cite{Mi3}.
Other examples of higher dimensions but of rank~2 include SO(5), Spin(5) and Sp(2).
In this case $2[\text{SO}(5), \ensuremath{\mathcal L}]=4 [\text{Sp}(2), \ensuremath{\mathcal L}]$, as shown in~\cite{Kn}.
In terms of generators~\cite{BS, Kn, Sm,Wo} one has $[\text{Sp}(2), \ensuremath{\mathcal L}]=\pm \beta_1 \in
\pi_{10}^s\cong {\mathbb Z}_6$.
On the other hand, for SO(4) the class is $[\text{SO}(4), \ensuremath{\mathcal L}]=0$.
For the case of the projective group, since $\text{SO}(5)\cong \text{PSp}(2)$ then $2[\text{PSp}(2), \ensuremath{\mathcal
L}]=4[\text{Sp}(2), \ensuremath{\mathcal L}]$.

 {\bf 4.~Stiefel manifolds.} For $1 \leq q \leq n-1$, let $V_{n,q}$ denote the Stiefel manifold of orthogonal $q$-frames in
$\mathbb{F}^n$, where $\mathbb{F}={\mathbb R}, {\mathbb C}, {\mathbb H}$.
Then, from~\cite{Mi}, the element $[V_{n,q}~, \phi ]=0$ for a~framing $\phi$ in the sense of~\cite{LS}.

  {\bf 5.~Flag manifolds.}
Let $\phi$ be a~stable framing of the framed f\/lag manifold $G/T$ and $[G/T, \phi]\in \pi_*^s$.
Then $2[G/T, \phi]=0 \in \pi_*^s$.
This implies that there is a~framing $\phi$ of the eight-dimensional f\/lag manifold $\text{Sp}(2)/T^2$ such that $ [{\rm
Sp}(2)/T^2;\phi]=\; \eta\circ
\sigma \in \pi_8^s$, where $\eta \neq 0\in \pi_1^s$, as above.

We now highlight a~few useful properties of Lie groups elements in framed cobordism.

{\bf Properties.} {\it $1$.~Product of groups~{\rm \cite{BS}}:} For two groups $G$ and $H$ with framings $\ensuremath{\mathcal L}_G$ and
$\ensuremath{\mathcal L}_H$, respectively, the corresponding classes satisfy $[G, \ensuremath{\mathcal L}_G] \times [H,
\ensuremath{\mathcal L}_H]=[G \times H, \ensuremath{\mathcal L}_{G \times H}]$.
This implies, for example, that $T^2$, $T^3$, $S^3 \times S^3$, and $S^3 \times S^3 \times S^3$ with their
left-invariant framings give nonzero elements in $\pi_*^s$.

 {\it $2$.~Effect of change of orientation~{\rm \cite{ASm}}:} Changing the orientation results in a~possible reversal of sign of the
corresponding class $[G, \alpha, \ensuremath{\mathcal R}]= (-1)^{\dim G}[G, -\alpha, \ensuremath{\mathcal L}]=(-1)^{\dim
G + 1}[G, \alpha, \ensuremath{\mathcal L}]$.

 {\it $3$.~Effect of change of representation~{\rm \cite{Mi2}}:} For any two real representations $\rho_1$ and~$\rho_2$ of~$G$, and for
the adjoint representation Ad${}_G$, the following holds $ [G, \text{Ad}_G-\rho_1 + \rho_2]=(-1)^{\dim G}[G,
\rho_1-\rho_2]$.

The consequences of the above properties can be summarized in that the only ef\/fect at the quantum theory is a~possible
sign change in the cobordism invariants discussed below.
Therefore, if the ef\/fective action is already integral then a~change in sign would not af\/fect the single-valuedness of
the function.

{\bf Relation between framed cobordism and Spin cobordism.} In positive dimensions, the image of $\Omega^\text{fr}_* \to
\Omega_*^\text{Spin}$ is zero unless $*=8k+1$ or $8k+2$, where it is~${\mathbb Z}_2$ and is detected by the
Atiyah-Milnor-Singer $\alpha$-invariant.
See~\cite{DMW-Spinc} for an extensive discussion on the applications to string theory in ten dimensions.
Since $\Omega_{11}^\text{Spin}=0$, the map $\pi^s_{11}\to \Omega_{11}^\text{Spin}$ is trivial and so a~framed
eleven-dimensional manifold $Y^{11}$ may be viewed as the boundary of a~Spin mani\-fold~$Z^{12}$ of dimension $12$ with
the induced Spin structure on $Y^{11}$ being compatible with the framing.

\subsection{Framed cobordism invariants at chromatic level 1}

We will describe how cobordism invariants of framed manifolds appear in the description of the partition function in
M-theory and type II string theory.
To that end we f\/irst describe these invariants within framed cobordism.
In particular, we will describe how the $d$-invariant and the $e$-invariants appear, thus implementing some of the
entries appearing in the table in the Introduction.

\subsubsection[The $d$-invariant and Arf invariant in type II string theory]{The $\boldsymbol{d}$-invariant and Arf invariant in type II string theory}
\label{Sec d}

Here we recall two invariants relevant for the partition function in dimension ten (and to some extent in dimension
nine).
These two invariants are in fact very closely related.

{\bf 1.~The Arf invariant.} Algebraically, Arf invariants are def\/ined for quadratic forms over f\/ield of characteristic two and,
therefore, occur in various guises.
Their most prominent occurrence in topology is that of the Arf-Kervaire invariant of a~framed manifold, which is
a~framed cobordism invariant def\/ined in dimensions $4k+2$, $k\geq 0$,
\begin{gather*}
\text{Arf}: \Omega_{4k+2}^\text{fr} \to {\mathbb Z}_2.
\end{gather*}
While this invariant vanishes on closed, framed 10-manifolds, a~variant of this invariant is used in the construction of
the partition function~\cite{DMW, MoS, DMW-Spinc}.
The importance of such a~variant for type IIB string theory, as well as for the M5-brane, is highlighted in~\cite{BM,
IIB-global}.
The Arf-Kevaire invariant is, however, relevant on 6-manifolds.
A mathematical discussion on the relation to the M5-brane anomaly (involving manifolds with corners) can be found
in~\cite{HS}.
This is further amplif\/ied in~\cite{M2M5-framing}.

{\bf 2.~The $\boldsymbol{d}$-invariant.} Using K-theory, Adams def\/ined surjective homomorphisms, the $d$-invariants
\begin{gather*}
d_{\mathbb R}: \ \Omega_9^\text{fr} \cong \pi_{9}S^0 \to {\mathbb Z}_2,
\qquad
\Omega_{10}^\text{fr} \cong \pi_{10}S^0 \to {\mathbb Z}_2.
\end{gather*}
These are given by the mod~2 index of the Dirac operator~\cite{AS-skew, AS5}.
An extensive discussion in the context of M-theory can be found in~\cite{DMW-Spinc}.

{\bf Examples.
1.~Lie groups.} The $d$-invariant $d_{\mathbb R}: \pi_n^s \to {\mathbb Z}_2$, for $n\equiv 1$ or 2 mod~8, vanishes for any
non-abelian compact Lie group $G$.
In fact, on such a~group (e.g.\
U(3) or SO(5)) there is a~bi-invariant metric of positive scalar curvature.
Hence, by the Lichnerowicz theorem, there are no harmonic spinors on $G$.
But from~\cite{AS4}, $d_{\mathbb R}$ is the real (resp.\
complex) dimension mod~2 of the space of harmonic spinors on~$G$.
Thus, $d_{\mathbb R}[G]=0$.
In fact, if~$G$ is a~compact Lie group then $d_{\mathbb R}([G, \ensuremath{\mathcal L}])=0$ except in low
dimensions~\cite{ASm}.

  {\bf 2.~Finite quotients of Lie groups.} Let $G$ be a~semisimple Lie group of dimension ten with non-abelian maximal compact
subgroup, for example the Lorentz group SO(1,4).
Let $\Gamma$ is a~discrete subgroup such that $G/\Gamma$ is compact.
Then $d[G/\Gamma]=0$~\cite{Si1}.
The same holds for $G$ a~simply connected nilpotent Lie group.
Now, from Atiyah--Singer index theorem, the $d$-invariant is given by the kernel $\operatorname{Ker} D$ of the Dirac operator $D$ on
$G/\Gamma$
\begin{gather*}
d[G/\Gamma]=h(G/\Gamma)=\tfrac{1}{2}\dim (\operatorname{Ker}D).
\end{gather*}
It turns out that $h(G/\Gamma)$ is an even integer so that, in the context of~\cite{DMW, DMW-Spinc}, the partition
function is anomaly free.

Note that the result does not apply to general parallelizable manifolds.
A counterexample, kindly provided by one of the referees, is the following.
By surjectivity of the degree $d_{\mathbb R}$, there is a~closed framed 9-manifold $M^9$ with nontrivial Dirac index
modulo two.
The circle, equipped with its non-bounding framing, has a~nontrivial degree as well.
Consequently, the same holds true for the product $M^9 \times S^1$ (with the product framing), which is certainly
parallelizable.

\subsubsection[The $e$-invariant and M-theory]{The $\boldsymbol{e}$-invariant and M-theory}
\label{Sec e}

Using K-theory, Adams def\/ined surjective homomorphisms, the $e$-invariant\footnote{Note
that there is a~third variant of the Adams $e$-invariant, viz.
$e: \ker (d_{\mathbb R}: \pi_{8k+1}^s S^0 \to {\mathbb Z}_2) \to {\mathbb Z}_2$, detecting the image of the
$J$-homomorphism in these dimensions.},
\begin{gather}
e:~\pi_{4k-1}S^0 \to {\mathbb Z}_{d_k},
\qquad
\pi_{8k} S^0 \to {\mathbb Z}_2,
\label{ee}
\end{gather}
where $d_k$ denotes the denominator of $B_{2k}/4k$, where $B_i$ is the Bernoulli number.
These numbers are the orders of the corresponding cobordism groups, which are $\Omega_3^\text{fr}\cong {\mathbb Z}_{24}$,
$\Omega_7^\text{fr} \cong {\mathbb Z}_{240} \cong {\mathbb Z}_3 \oplus {\mathbb Z}_5 \oplus {\mathbb Z}_{16}$ (with
generator $S^7$ with twisted framing def\/ined by the generator of $\pi_7(O)\cong {\mathbb Z}$) and $\Omega_{11}^\text{fr}
\cong {\mathbb Z}_{504} \cong {\mathbb Z}_7 \oplus {\mathbb Z}_8 \oplus {\mathbb Z}_9$.
We are interested in $k=1, 2, 3$ in the f\/irst case and $k=1$ in the second case in~\eqref{ee}.
It is only a~low-dimensional `accident' that the real $e$-invariant can detect all of $\pi_{4k-1}^s S^0$ for $1 \leq k
\leq 3$.
In general, ${\mathbb Z}_{d_k}$ occur as direct summands, i.e.\
the cockerel of the $J$-homomorphism in dimension $4k-1$ is usually nontrivial.

{\bf The $\boldsymbol{e}$-invariant.} A $U$-structure is a~lift (up to homotopy) of the classifying map of the tangent bundle
$TX^\text{st}: X \to \text{BO}$ to $\text{BU}$.
A $(U,\text{fr})$-manifold is a~compact $U$-manifold $X$ with smooth boundary and a~trivialization of $E \cong TX^{\rm
st}$ over the boundary, i.e.\
a~bundle map $\psi: E|_{\partial X} \cong \partial X \times {\mathbb C}^k$.
In particular, $\psi$ provides a~framing for $\partial X$.
Using relative characteristic classes of the complex vector bundle $E \cong TX^\text{st}$, the {\it complex
$e$-invariant} of the framed bordism class of $\partial X$ is def\/ined to be
\begin{gather*}
e_{\mathbb C} (\partial X) \equiv \langle \text{Td}(E), [X, \partial X] \rangle
\mod {\mathbb Z}.
\end{gather*}
By Atiyah--Patodi--Singer~\cite{APS} the quantity on the right hand side is $\langle \text{Td}(E),\! [X, \partial X]\rangle
{=} \! \int_X \! \text{Td} (\nabla^E),$ where $\nabla^E$ is a~unitary connection on $E$ which restricts to the canoncial
f\/lat connection specif\/ied by the trivialization.
Similarly, the real $e$-invariant $e_{\mathbb R}: \pi_{4k-1}^\text{S} \to {\mathbb Q}/{\mathbb Z}$ can be def\/ined for
Spin manifolds.
The two are related by $e_{\mathbb R}/\epsilon \equiv e_{\mathbb C}\mod{\mathbb Z}$, where $\epsilon (k)=1$ if $k$
even and $\tfrac{1}{2}$ otherwise.
See~\cite{tcu, DMW-Spinc} for applications to M-theory.
In that context, since $TZ^{12}$ is trivialized over~$Y^{11}$ we can def\/ine the relative Pontrjagin classes $p_i$ in
$H^{4*}(Z^{12},Y^{11})$ and hence evaluate the $\widehat{A}_k$-polynomial on the fundamental cycle of $Z^{12}$.
Then the $e$-invariant of $Y^{11}$ can be def\/ined via the relative $\widehat{A}$-genus as~\cite{ASm}
\begin{gather*}
e\big[Y^{11}\big]=\tfrac{1}{2}\widehat{A}\big(Z^{12}\big) \mod {\mathbb Z}.
\end{gather*}
By the index theorem, this is independent of the choice of the bounding manifold $Z^{12}$.
This can be viewed as an analog of the similar observation on the ef\/fective action in~\cite{DMW} for the Spin case.

{\bf Examples.} {\bf 1.~Spheres.} Consider those spheres that are Lie groups.
In this case, $e[S^1]=\pm \frac{1}{2}$ and $e[S^3]=\pm \frac{1}{24}$ (see~\cite{ASm}).
For applications to the M2-brane see~\cite{tcu}.

  {\bf 2.~Tori.} The 3-dimensional torus $T^3=S^1 \times S^1 \times S^1$ has $e$-invariant $\frac{1}{2}$.
This cannot be directly generalized to higher dimensional tori, due to nil potency, i.e.\
$\eta^4=0$.
Therefore, any torus of dimension greater than three represents zero when equipped with the left-invariant framing.

 {\bf 3.~Quotients or extensions.} For $C$ a~central subgroup, knowing the $e$-invariant of the Lie group $G$ allows us to know
that of the quotient $G/C$ and vice-versa.
For example, the relation $[\text{SO}(3), \ensuremath{\mathcal L}]=2[\text{Spin}(3), \ensuremath{\mathcal L}]=2[S^3,
\ensuremath{\mathcal L}]$ gives $e[\text{SO}(3)]=\pm \frac{1}{12}$.

  {\bf 4.~SU(3)}.
The group SU(3) as a~framed manifold with a~left-invariant framing represents $[\text{SU}(3), \ensuremath{\mathcal
L}]=\overline{\nu}\in \pi_8^s={\mathbb Z}_2 \oplus {\mathbb Z}_2\cdot \overline{\nu}$~\cite{St, Wo}.
Let $\lambda: \text{SU}(3) \to \text{SU}(3)$ denote the identity map regarded as the fundamental representation of SU(3)
on ${\mathbb C}^3$.
Then the $e$-invariant of $[\text{SU}(3), \lambda]$ is nonzero~\cite{Wo}.

  {\bf 5.~U(3).} Let $
\sigma$ be the generator of $\pi_2^s(BS^1)$ given by the Hopf bundle.
Then $t(\sigma)=[S^3, \ensuremath{\mathcal L}]$ generates $\pi_3^s(S^0)$, so that $t(3\sigma)=\nu$.
From $\eta \overline{\nu}
=\nu^3$, this gives $[\text{U}(3), \ensuremath{\mathcal L}]=\eta [\text{SU}(3), \ensuremath{\mathcal L}]=\nu^3$.

{\bf Integrality and corners.} The Todd genus of a~$(U,\text{fr})$-manifold is integral if and only if the complex
$e$-invariant $e_{\mathbb C}$ of its boundary is integral or, equivalently, if and only if its boundary is the corner
$\partial_0 \partial_1 X$ of a~$(U, \text{fr})^2$-manifold $X$.
In the context of M-theory, this says that the index~-- in the form of the Todd genus~-- of $Z^{12}$ is integral if and
only if $Y^{11}=\partial_0 \partial_1 W^{13}$ of a~$(U, \text{fr})^2$-manifold $W^{13}$.
This views $Y^{11}$ itself as a~corner, in contrast to viewing its boundary as a~corner, as we do for most of this
article.
The structure of the topological terms in M-theory indeed do not suggest a~lifting to thirteen dimensions.

\subsection{Change of framing}
\label{Sec ch}

Like other geometric structures on manifolds (e.g.\
Spin structure), a~framed manifold might admit dif\/ferent framings.
In this section we will study the possible physical ef\/fect of the change of framing.
The analog for Spin and Spin${}^c$ structures is studied in~\cite{DMW-Spinc}, and that of String structures
in~\cite{tcu}.

A framing on a~manifold $M$ corresponds to a~lift in the diagram
\begin{gather*}
\xymatrix{ && \text{EO}(n) \ar[d] & \text{O}(n) \ar[l]
\\
M^n \ar[urr] \ar[rr]^\nu && \text{BO}(n)
}
\end{gather*}
The dif\/ferent choices of framing correspond to maps from $M^n$ to the f\/iber $\text{O}(n)$ of the principal classifying bundle.
Given a~framed manifold $M^n$ and a~map $\mathcal{F}: M^n \to \text{O}(n)$ we may use $\mathcal{F}$ to change the framing.
Conversely, given two framings $\phi_1$ and $\phi_2$ of $M$, they dif\/fer by a~map $\phi_2/\phi_1: M^n \to \text{O}(n)$.
The change of framing in the case of two and ten dimensions can be viewed from the point of view of the Atiyah
$\alpha$-invariant~\cite{tcu, DMW-Spinc} which is the ref\/inement of the mod~2 index of the Dirac operator from ${\mathbb
Z}_2$ to $\text{KO}_m\cong{\mathbb Z}_2$ for $m=2, 10$.

{\bf Twisted framing.} We start with the case of a~Lie group, which is always oriented as a~manifold.
Given a~map $\varphi: G \to \text{SO}(n)$, there is an automorphism of the trivial bundle $G \times {\mathbb R}^d$ given by
$(g, {\omega})\mapsto (g, \varphi{g}^{-1}(w))$.
Then the twisted framing (here for right) $\ensuremath{\mathcal R}^\varphi$ of $\ensuremath{\mathcal R}$ by $\varphi$ is
def\/ined as the direct sum of $\ensuremath{\mathcal R}$ with this automorphism.
The determination of $[G, \ensuremath{\mathcal R}^\varphi]$ depends essentially on the element of the reduced group
$\widetilde{\text{KO}}^{-1}(G^+)$ represented by $\varphi$, where $G^+$ is $G$ with a~point adjoined.
Given an element $\alpha$ in $\widetilde{\text{KO}}^{-1}(G)$ we may twist a~given framing $\phi$ of $G$ to obtain a~new element
$[G, \phi^a]$.
Let $M^n$ be a~framed manifold embedded in ${\mathbb R}^{n+k}$ and $\phi$ a~framing of its normal bundle $NM$.
For $\alpha\in \widetilde{\text{KO}}^{-1}(M)$ the twisted framing $\phi^\alpha$ is constructed as the composition of $\phi$ and
the automorphism $\tilde{\alpha}$ of the trivial bundle $M \times {\mathbb R}^k$ determined by $\alpha$.

{\bf Example.
Change of framing on sphere bundles over spheres.} We consider the example we discussed earlier in Sections~\ref{Sec par}
and~\ref{Sec st}.
The framing $\phi$ of $S(E)$ is called the induced framing.
All other framings of $S(E)$ may be obtained from $\phi$ by twisting with elements of $\widetilde{\text{KO}}^{-1}(S(E))$.
Suppose that $[E]=0\in \widetilde{\text{KO}}(S^n)$.
Then $S(E)$ is a~$\pi$-manifold and hence there are elements $ [S(E); \phi] \in \pi_{n+m-1}^s $ corresponding to the
dif\/ferent framings $\phi$ of $S(E)$.

{\bf Change of framing and the $\boldsymbol{d}$-invariant.}  The change of framing will have an ef\/fect on the $d$-invariant.
We consider the case of a~Lie group, i.e.\
a~WZW model context.
For an element~$\alpha$, giving rise to a~left framing $\ensuremath{\mathcal L}^\alpha$, the $d$-invariant will get
modif\/ied by~$J(\alpha)$ as
\begin{gather*}
d_{\mathbb R}([G, \ensuremath{\mathcal L}^\alpha])=d_{\mathbb R} (J(\alpha)) \circ d_{\mathbb R}([G,
\ensuremath{\mathcal L}]).
\end{gather*}
In order to guarantee the absence of mod~2 anomalies in the partition function, we would like the transformed
$d$-invariant to be even.
If $d_{{\mathbb R}}([G, {\ensuremath{\mathcal L}}])$ started out as being already even then there are no conditions
needed.
However, if $d_{{\mathbb R}}([G, \ensuremath{\mathcal L}])$ were odd then there are no potential anomalies if
$d_{\mathbb R}(J(\alpha))$ is also even.
If this occurs, then the change of framing could be viewed as a~way of curing an anomaly.

Note that, due to the index interpretation, questions about the degree can be rephrased in terms of Spin geometry
(see~\cite{DMW-Spinc} for extensive discussion in the context of M-theory).
In particular, for WZW models on compact non-abelian groups, the scalar curvature argument ensures triviality of the
degree (independent of the framing/Spin structure).
On the abelian groups $S^1$ and $S^1 \times S^1$, however, it is easy to write down reframings which change the Spin
structures and the degree.
Unfortunately, these targets do not give rise to `honest' WZW models.

{\bf Change of framing and the $\boldsymbol{e}$-invariant.}  Let $s^*: H^*(\text{BSO}; {\mathbb Q}) \to H^*(\text{SO};{\mathbb Q})$ be the
cohomology suspension and set $u_n=s^*p_n \in H^{4n-1}(\text{SO};{\mathbb Q})$ where $p_n \in H^{4n}(\text{BSO};{\mathbb Q})$ is the
$n$th universal Pontrjagin class.
The cohomology suspension kills decomposable elements\footnote{This is a~process used in~\cite{S-Sig}.} so that
\begin{gather*}
s^*\widehat{A}_n(p)=-\frac{B_n}{4n}\frac{s^*p_n}{(2n-1)!}= -\frac{B_n}{4n}\frac{u_n}{(2n-1)!},
\end{gather*}
where $B_n$ is the $n$th Bernoulli number.
Suppose that $E \to S^n$ is stably trivial $m$-plane bundle, with $\phi$ a~stable trivialization of $E$, and $\alpha \in
\widetilde{\text{KO}}^{-1}(S(E))$.
Then putting $4k=n+m$, the $e$-invariant with new element is~\cite{Sm}
\begin{gather*}
e[S(E); \phi_\alpha]=-\frac{a_kB_k}{4k(2k-1)!} \left\langle \alpha^* u_k, [S(E)] \right\rangle.
\end{gather*}
Since Lie groups are admissible manifolds which, furthermore, do not lead to anomalies in the partition function, we
should have that $\alpha^* u_3$ is an even multiple of $u_3$ for an 11-dimensional bundle with a~spherical base.

{\bf Examples.}  {\bf 1.}~Let $G$ be a~compact connected Lie group of dimension $3$ and $\alpha \in
\widetilde{\text{KO}}^{-1}(G)$ be an element with second Stiefel--Whitney class $w_2(\alpha)$ zero.
If $p_1(\alpha)$ is the f\/irst Pontragin class of $\alpha$ in $H^{3}(G;{\mathbb Z})$ (i.e.\
via transgression to the Chern--Simons form) then the $e$-invariant of the variation is given by~\cite{Kn}
\begin{gather}
e_{\mathbb R}\big([G,\ensuremath{\mathcal L}^\alpha] - [G,\ensuremath{\mathcal L}]\big)
= - \left\langle \tfrac{1}{4}B_{2}\cdot p_1(\alpha),[G]\right\rangle_H\mod 2{\mathbb Z}.
\label{eee}
\end{gather}
From the results in~\cite{DMW-Spinc}, we require the cohomological pairing on the right hand side of~\eqref{eee} to be
an integer.
This places an obvious congruence condition on the Pontrjagin class $p_1(\alpha)$.
The groups we have in mind are $\text{SU}(2)\cong \text{Spin}(3) \cong \text{Sp}(1)$ and SO(3).

  {\bf 2.}~If $G$ is a~compact connected Lie group of dimension $m=$8 or 9, and $\lambda \in
\widetilde{\text{KO}}^{-1}(G)$ satisf\/ies $w_2(\lambda)=0$ then~\cite{Wo}
\begin{gather}
e_{\mathbb R}\big([G, \ensuremath{\mathcal L}^\lambda]- [G, \ensuremath{\mathcal L}]\big)
=\big\langle \rho^3(E_\lambda)-1,[G]\big\rangle_{{}_{\text{KO}}} \in \text{KO}_{m+1}(\ast)\cong{\mathbb Z}_2.
\label{KO}
\end{gather}
Here $\rho^3$ is the cannibalistic characteristic class of Adams and Bott associated with the Thom isomorphism
$\rho^3(E)=\phi_E^{-1} \circ \psi^3 \circ \phi_E(1)$.
The groups we have in mind here are SU(3) in dimension~8 and~U(3) in dimension~9.
The Adams operation $\psi^i$, which is an automorphism of K-theory, is given a~physical interpretation in the context of
M-theory in~\cite{DMW-Spinc}.
As in the previous case, the results of~\cite{DMW-Spinc} require that the KO-theoretic pairing on the right hand side
of~\eqref{KO} to be zero in~${\mathbb Z}_2$.

The consequence of the above two examples is that the dif\/ference of the $e$-invariant should be an even integer in order
for the partition function to be anomaly-free.

\section{Topological and modular aspects of M-theory with corners}

\subsection{Building the general setting in M-theory}
\label{Sec build}

We will show in this section how the formulation in terms of corners is the natural setting for topological
considerations of M-theory.
This complements the analytical point of view in~\cite{DMW-corner}.

Consider M-theory on an 11-dimensional manifold $Y^{11}$ with boundary $M^{10}=\partial Y^{11}$.
This is considered in~\cite{DFM}, where
a~model of the $C$-f\/ield which is valid in that case is proposed, with the phase given by
\begin{gather*}
\Phi \big(C, Y^{11}\big)= \exp \left[ i\pi \xi (D_A) + \tfrac{i\pi}{2} \xi (D_{\rm RS}) + 2\pi i I_\text{local} \right],
\end{gather*}
where $\xi(D_A)$ is a~section of a~U(1) bundle with connection over the space of $C$-f\/ields on the boundary $M^{10}$, with
$D_A$ an $E_8$ Dirac operator and $D_{\rm RS}$ the Rarita--Schwinger operator, and $\xi$ denotes the reduced eta-invariant
$\xi(D)=\frac{1}{2}(\eta (D) + h(D))$.
The model is also extended to the case when the 11-manifold is of the form $Y^{11}=X^{10}\times S^1$ with the type IIA
10-manifold $X^{10}$ itself having a~boundary $N^9=\partial X^{10}$, so that $M^{10}= N^9 \times S^1$.

The authors of~\cite{DFM} also consider the reduction of the phase from $Z^{12}$ to $M^{10}$ in the product
case\footnote{They consider $Z^{12}= M^{10} \times \Delta^2$, where $\Delta^2$ is a~2-simplex $\{ (t_1, t_2): 0 \leq t_1
\leq t_2 \leq 1 \}$ with identif\/ication $t_i
\sim t_i +1$.}.
With the ansatz for the f\/ield strength $G_Z= G_M + dt_1 \wedge \omega_1 + dt_2 \wedge \omega_2$, the phase reduced to
$M^{10}$ takes the form $\exp \left[ -2\pi i \frac{1}{2}\int_{M^{10}} G \wedge \omega_1 \wedge \omega_2 \right]$.
We will propose a~generalization of this, in some sense, to manifolds with corners.
The authors consider conditions on torsion cohomology classes; we will in addition be interested in doing global
analysis.
As the structure above suggests, we will highlight the case when the corner $M^{10}$ is the total space of a~circle
bundle $\pi$.
The pairing $H^7(N^9; {\mathbb Z}) \times H^2(N^9; U(1)) \to U(1)$ suggests that the structure of the base of that
circle bundle $\pi$ has a~7-dimensional factor.
Indeed, the most interesting case will be when $M^{10}=S^3 \times S^7$, corresponding to a~base $N^9= S^7 \times S^2$ or
$N^9={\mathbb C} P^3 \times S^3$.

We generalize the above setting of product structure to more general bundle structure captured by this schematic diagram
\begin{gather}
\begin{split}
& \xymatrix{ & & Z^{12} \ar[dl]_{\partial_1} \ar[dr]^{\partial_2} & &
\\
{\mathbb{D}}^2 \ar[r] & W^{11}=\partial_1 Z^{12} \ar[dr]^\partial \ar[dd] & & {\underbrace{\partial_2
Z^{12}=Y^{11}}_\text{M-theory}} \ar[dd] \ar[dl]_{\partial_2} & S^1 \ar[l]
\\
& & {\underbrace{M^{10}}_\text{Heterotic}} \ar[dl]_\pi & &
\\
& N^9=\partial X^{10} & & {\underbrace{X^{10}}_\text{type~IIA}} \ar[ll]_\partial & U^{11}
\ar[l]_{~~~~\partial} }
\end{split}
\label{Diag}
\end{gather}

The extension of type IIA to $U^{11}$ is discussed in~\cite{BM, S-gerbe}.
General boundary conditions on the $C$-f\/ield (and its dual) are studied in~\cite{DMW-boundary}.

{\bf The Rarita--Schwinger f\/ield and supersymmetry.}
\label{Sec RS}
We highlight some ef\/fect on the spinor f\/ields and, in particular, as far as stable trivialization is involved.
The Rarita--Schwinger operator over a~manifold $M$ is def\/ined as the Dirac operator twisted by the virtual bundle $RS=TM
\ominus m \mathcal{O}$, for some multiple of the trivial line bundle $\mathcal{O}$.
For example, $m=4, 3, 2$ in twelve, eleven, and ten dimensions, respectively.
Had the twisting bundle $RS$ not been a~virtual bundle, it would have been stably trivial  if $RS \oplus n{\cal O}$ is
trivial.
This then would imply that $TM$ is trivial, and hence that $M$ is parallelizable (see Section~\ref{Sec par}).
However, $RS$ is only a~virtual bundle and, as such, triviality only makes sense in $\widetilde{\text{KO}}(M)$ and, in
particular, such a~triviality cannot be used to deduce strict parallelizability of $M$.
Of course, if $TM$ is trivial then $RS$ would be.
In this case, the characteristic classes of the tangent (and Spin) bundle will be trivial.
Triviality of the tangent bundle implies the triviality of the Spin bundle.
Parallelizability implies the maximum number of linearly independent sections.
Thus, for the Spin bundle this ensures the maximum number of spinors.
This is desirable for compactif\/ication to low dimensions.

\subsection{Relation to type IIA: Disk bundles and eta-forms}

In this section we will consider the ef\/fective action and partition function of type IIA string theory in the context of
corners.
We will identify the contribution to the phase.
This will allow us to provide physical interpretation of higher eta-forms, extending the discussion in~\cite{MaS,S-gerbe}.

Let $Z$ be the total space of a~disk bundle over a~$\pi$-manifold $X$ associated with a~vector bund\-le~$E$.
For the purpose of relating M-theory to type IIA string theory, $E$ is usually a~hermitian line bundle.
The Kervaire semi-characteristic is zero, so that if the circle bundle $Y^{11}$ is stably parallelizable then it is also
parallelizable.

{\bf $\boldsymbol{Z^{12}}$ as a~$\boldsymbol{\langle 2 \rangle_f}$-manifold.} We consider $Z^{12}$ as a~f\/iber bundle $\pi: Z^{12} \to X^{10}$,
where both the disk f\/iber $\mathbb{D}^2$ and the type IIA base $X^{10}$ are compact $\langle 1 \rangle$-manifolds, i.e.\
manifolds with boundary, and the faces are given by the f\/iber bundles (see diagram~\eqref{Diag})
\begin{gather}
\partial_1 Z^{12}= W^{11}:
\quad
\mathbb{D}^2 \to W^{11} \to \partial X^{10}=N^9,
\nonumber
\\
\partial_2 Z^{12}=Y^{11}:
\quad
S^1=\partial \mathbb{D}^2 \to Y^{11} \to X^{10}.
\label{second}
\end{gather}
Consider metrics $g^{TX}$ and $g^{TD}$ and connections $\nabla^{TX}$ and $\nabla^{TD}$ on $TX^{10}$ and $T\mathbb{D}^2$,
respectively.
Corresponding to the splitting $TZ^{12} \cong T\mathbb{D}^2 \oplus \pi^* TX^{10}$, we have a~connection
$\nabla^\oplus=\nabla^{TD} \oplus \pi^* \nabla^{TX}$.
Consider a~Dirac family $D_E$ on $Z^{12}$ parametrized by points on the base, as in~\cite{MaS, S-gerbe, S-Sig,
DMW-boundary, DMW-corner}, assuming a~metric of product type near the boundary.
Furthermore, we assume that the kernel of $D_E$ induced on the f\/iberwise boundary~\eqref{second} is of constant rank, so
that the index bundle with respect to the Atiyah--Patodi--Singer boundary problem is well-def\/ined, with $D_E^\partial$
the corresponding twisted Dirac operator on the boundary.
We also consider the line bundle $L$ with connection $\nabla^L$ corresponding to the principal circle bundle
(see~\cite{DMW-Spinc}).

{\bf Contribution to the phase from the boundary of type IIA.} The Bismut--Cheeger~\cite{BC} and Melrose--Piazza~\cite{MP}
formulation of the index implies that a~representative in cohomology of the Chern character of the index bundle is given
by the following dif\/ferential form on $X^{10}$:
\begin{gather*}
\int_{\mathbb{D}^2} \left\{ \widehat{A}\big(\nabla^{TD}\big) \operatorname{ch}\big(\nabla^{L^{1/2}}\big) \operatorname{ch}\big(\nabla^V\big) \right\} -
\widehat{\eta}(D^\partial_V) -\tfrac{1}{2} \operatorname{ch} \big(\nabla^{\ker D^\partial_V} \big).
\end{gather*}
The Index class can be encoded in a~virtual vector bundle $\xi$ with unitary connections $\nabla^\xi$, such that
$[\xi]=[\xi_1 \ominus \xi_2]=[\text{Ind}]$, so that in de Rham cohomology $\left( \operatorname{ch}(\text{Ind})\right)_{\rm
dR}=\operatorname{ch} (\nabla^\xi) + d\omega$, for some $\omega \in \Omega^\text{odd}(X^{10})$.
We now introduce the $E_8$ bundle $V$ with connection $\nabla^V$ on $Z^{12}$ and its restriction to the boundary, with
the same notation, and assuming a~pull-back structure, i.e.\
with the bundle $E$ pulled back fro the base.
This is a~reasonable assumption from the physical point of view, as in~\cite{DMW, MaS, S-gerbe}.
Furthermore, we assume that the Dirac family on the (f\/iberwise) boundary should be
invertible\footnote{Alternatively, one could absorb additional corrections by redef\/ining the $\widehat{\eta}$-forms.}. Then, with
a~product connection $\nabla^\oplus$, via~\cite{Bod}
\begin{gather}
\int_{Z^{12}} \widehat{A}\big(\nabla^\oplus\big) \operatorname{ch}\big(\pi^* \nabla^V\big) = \int_{Z^{12}} \widehat{A} \big(\pi^*
\nabla^{TX}\big) \widehat{A}\big(\nabla^{TD}\big) \operatorname{ch}\big(\pi^* \nabla^V\big)
\nonumber
\\
\qquad
=\int_{X^{10}}\left\{\widehat{A}\big(\nabla^{TX}\big) \operatorname{ch}\big(\nabla^V\big)
\big(\operatorname{ch}\big(\nabla^\xi\big)+d\omega+\widehat{\eta}\big)\right\}
\label{big}
\\
\qquad
= \text{Ind} \big(D_{\xi \otimes V}\big) +\!\int_{X^{10}}\! \widehat{A}\big(\nabla^{TX}\big) \operatorname{ch}\big(\nabla^V\big) \widehat{\eta} +
\left\{
\xi \big(D^{\partial X}_{\xi \otimes V}\big) + \!\int_{\partial X^{10}}\! \omega \widehat{A}\big(\nabla^{TX}\big) \operatorname{ch} \big(\nabla^V\big)
\right\} .\nonumber
\end{gather}
The phase of the partition function is $\text{Phase}=\text{Ind}(D_{\xi \otimes V})$.
The last term in~\eqref{big} is the contribution from the boundary in type IIA.

{\bf Interpretation of the eta-forms.} The basic part of expression~\eqref{big} gives
\begin{gather*}
\int_{\mathbb{D}^2} \widehat{A}(\nabla^{TD})=\left( \operatorname{ch}(\text{Ind})\right)_\text{dR} + \widehat{\eta}.
\end{gather*}
For the three relevant nontrivial degrees we have
\begin{gather*}
\int_{\mathbb{D}^2} \widehat{A}_4(\nabla^{TD})=\left( \operatorname{ch}_1(\text{Ind})\right)_\text{dR} + a B_2,
\\
\int_{\mathbb{D}^2} \widehat{A}_8(\nabla^{TD})=\left( \operatorname{ch}_3(\text{Ind})\right)_\text{dR} + b B_2^3,
\\
\int_{\mathbb{D}^2} \widehat{A}_{12}(\nabla^{TD})=\left( \operatorname{ch}_5(\text{Ind})\right)_\text{dR} + c B_2^5,
\end{gather*}
where we identify the components of the eta-form as powers of the $B$-f\/ield.
The numerical coef\/f\/icients $a$, $b$, $c$ can be read of\/f for a~certain class of situations; for instance, with a~suitable
identif\/ication of the $B$ and $F$ f\/ields and using expressions~\eqref{eta F} below.
This generalizes to higher degrees the interpretation in~\cite{MaS, S-gerbe} of the eta-form in degree two as
essentially the $B$-f\/ield.

We can approach this from an another angle which makes it a~bit more physically explicit.
Consider type IIA string theory on $U^8 \times T^2$ with metric $ds_X^2= ds^2_{T^2} + t^2 ds^2_U$.
When the $E_8$ bundle is trivial and with the Ramond--Ramond f\/ields $F_0=0=F_4$ then the phase in this case,
via~\cite{MoS}, is
\begin{gather*}
\Phi_1=\int_{X^{10}} \left[ -\tfrac{1}{15}(F_2)^5 + \tfrac{1}{3}(F_2)^3\cdot \tfrac{1}{24}p_1- \tfrac{1}{2}F_2 \cdot
\widehat{A}_8 \right].
\end{gather*}
Comparing with the second term in~\eqref{big}, we have the following interpretation of the eta-forms
\begin{gather}
\widehat{\eta}_2=-\tfrac{1}{2}F_2,
\qquad
\widehat{\eta}_6=-\tfrac{1}{3}(F_2)^3,
\qquad
\widehat{\eta}_{10}=-\tfrac{1}{15}(F_2)^5.
\label{eta F}
\end{gather}
This is another interpretation of eta-forms in terms of the Ramond--Ramond 2-form rather than in terms of the $B$-f\/ield.
However, there is no conf\/lict as the two f\/ields can be identif\/ied for some topological purposes; see~\cite{NS5,DMW-Spinc}.

\subsection{The heterotic theory as a~corner}
\label{Sec as}

In this section we will provide another argument for why the corner setting is natural to consider for heterotic string
theory.
This complements the discussion above in Section~\ref{Sec build} as well as the analytic arguments in~\cite{DMW-corner}.
In particular, the structure of the heterotic anomaly points in a~natural way to a~corner formulation.
The heterotic theory, like other f\/lavors of string theory, will mostly arise on manifolds which are decomposable as
products or as f\/iber bundles, as in~\cite{DMW-boundary}.

{\bf The heterotic corner as a~framed cobordism class.} We will consider the heterotic corner as a~framed corner.
The invariant we will associate to this is a~framed cobordism invariant, and hence depends only on the framed cobordism
class.
Using the results of~\cite{BN}, the data that ref\/ines the corner $M^{10}$ into a~representative of a~framed cobordism
class
\begin{gather*}
[M^{10}] \in \Omega_{12}^{(U, \text{fr})^2},
\end{gather*}
and which interestingly matches the physical setting (see diagram~\eqref{Diag}), is the following
\begin{enumerate}\itemsep=0pt
\item
A decomposition $TM^{10} \cong T^0 M^{10} \oplus T^1 M^{10}$ of framed bundles.
This implements a~factorization, as e.g.\
dictated by anomaly cancellation (see below).

\item
Compact 11-dimensional manifolds $W^{11}$, $Y^{11}$ with boundary $\partial W^{11} \cong - \partial Y^{11}$.
The manifold $Y^{11}$ where M-theory resides, while $W^{11}$ is `physically hidden'.

\item
Decompositions $TW^{11} \cong T^0W^{11} \oplus T^1 W^{11}$ with complex structure on the f\/irst factor and framing on the
second factor; and $TY^{11}\cong T^0Y^{11} \oplus T^1 Y^{11}$ with framing on the f\/irst factor and complex structure on
the second factor.
For instance, in the product $Y^{11}= X^3 \times W^8$, we take $X^3$ to be framed and $W^8$ to be complex (e.g.\
a~Calabi--Yau manifold).

\item
The inclusions $M^{10} \hookrightarrow W^{11}$, $M^{10} \hookrightarrow Y^{11}$ identify
\begin{gather}
(T^1W^{11})\vert_{M^{10}} \cong T^1M^{10},
\qquad
(T^0W^{11})\vert_{M^{10}} \cong T^0M^{10},
\qquad
\text{as~framed~bundles},
\nonumber
\\
\label{Dec 1}
(T^0W^{11})\vert_{M^{10}} \cong T^0M^{10},
\qquad
(T^1W^{11})\vert_{M^{10}} \cong T^1M^{10},
\qquad
\text{as~complex~bundles}.
\end{gather}
\item
A manifold with corners $Z^{12}$ such that $\partial_0 Z^{12} \cong W^{11}$ and $\partial_1 Z^{12} \cong Y^{11}$.
This is the lift in diagram~\eqref{Diag}.

\item
A decomposition $TZ^{12} \cong T^0 Z^{12} \oplus T^1 Z^{12}$ of complex vector bundles such that

\item
The inclusions $W^{11} \hookrightarrow Z^{12}$ and $Y^{11} \hookrightarrow Z^{12}$ identify
\begin{gather}
T^0Z^{12}\vert_{W^{11}} \cong T^0W^{11},
\qquad
T^1 Z^{12}\vert_{W^{11}} \cong T^1 W^{11},
\nonumber
\\
T^0Z^{12}\vert_{Y^{11}} \cong T^0Y^{11},
\qquad
T^1 Z^{12}\vert_{Y^{11}} \cong T^1 Y^{11},
\label{Dec 4}
\end{gather}
as complex bundles.
Again, the identif\/ications~\eqref{Dec 1}, \eqref{Dec 4} are to be understood as identif\/ications of stable bundles.
\end{enumerate}
The 11-dimensional manifold $W^{11}$ is to satisfy the above conditions so that diagram~\eqref{Diag} is properly
implemented.

\subsubsection{Corners as the natural setting for heterotic anomalies}
\label{Sec an}
Consider the Green--Schwarz anomaly in the heterotic (and type I) theory.
The general formulation of anomalies requires considering a~degree twelve polynomial whose integration over the
10-dimensional manifold is the curvature of the anomaly line bundle.
Consider heterotic string theory on a~Spin manifold $(M^{10}, g)$ with Spin bundle $SM$ and with a~vector bundle $E$ of
structure group $E_8 \times E_8$ or Spin(32)$/{\mathbb Z}_2$, which captures the dynamics of the Yang--Mills f\/ields via
a~connection $A$ on $E$.
Then the $H$-f\/ield and its dual satisfy~\cite{Fr}
\begin{gather}
\tfrac{1}{2\pi} d\left( H_3u^{-2}+H_7u^{-4} \right) =  \left[ \text{ch}_2(A) - p_1(g)\right] u^{-2}
\nonumber\\
\qquad
{}+\left[\text{ch}_4(A) - \tfrac{1}{48} p_1(g)\text{ch}_2(A) + \tfrac{1}{64} p_1(g)^2 -
\tfrac{1}{48} p_2(g)\right]u^{-4},
\label{eq_Fr}
\end{gather}
where $u$ is the Bott generator.
The anomaly polynomial is given by
\begin{gather}
I_{12}=\int_{Z^{12}} X_4 \wedge X_8
\label{I12}
\end{gather}
with $H_3$ and $H_7$ providing the trivializations of the forms $X_4$ and $X_8$, respectively, as given in~\eqref{eq_Fr}.
The usual connection to ten dimensions is that one has to introduce a~local term in the action functional of the form
$\int_{M^{10}}B_2 \wedge X_8$.
Our interpretation of this is that it is implicit that we schematically have the following relations
\begin{gather*}
X_4= dH_3,
\qquad
H_3=dB_2,
\end{gather*}
which hints at a~direct relation between $X_4$ and $B_2$ had one been able to make sense of a~boundary of a~boundary.
With the lack of this, the formulation in terms of corners seems to be an alternative.
Note that one still cannot directly take a~boundary of a~boundary, neither in the homological or the cohomological
setting.
A proper interpretation at each step is given towards the end of Section~\ref{Sec fra}.
A similar argument holds for the dual f\/ield $H_7$ with the polynomial $X_8$.
Therefore, the structure of the anomaly and the process of anomaly cancellation have a~natural home in the setting of
manifolds with corners.
Such a~formulation also holds in other situations where topological anomaly cancellation occurs.
While we do not explicitly spell them out, it is obvious how our discussion could be adapted.

Note that the anomaly polynomial can be given in terms of the elliptic genus as~\cite{LNSW}
\begin{gather*}
I_{12}=\phi_\text{ell}\big|_{\text{12-form~coef\/f.~of~}q^0}.
\end{gather*}
Then a~natural question is what this corresponds to in ten dimensions.
Our formulation in terms of corners can also be viewed as providing an answer to this question.
Note that the two terms in the integrand in~\eqref{I12} are interpreted in terms of twisted string structure and twisted
Fivebrane structures, respectively~\cite{SSSII, SSSIII}.

\subsubsection{The one-loop term in the presence of a~corner}
\label{Sec 1}

The topological study of M-theory relies on the existence of the one-loop polynomial $I_8$~\cite{DLM}, which is a~degree
eight polynomial in the Pontrjagin classes of eleven-dimensional spacetime $Y^{11}$.
The term in the action is of the form $\int_{Y^{11}}C_3 \wedge I_8$.
It is natural to ask how this term behaves under dimensional reduction to ten or lower dimensions, or upon dimensional
lifting to twelve and higher dimensions.
An example of the former is in type IIA string theory, where the term takes the form $\int_{X^{10}}B_2 \wedge I_8$.
An example of the latter situation is the lift to the 12-dimensional bounding theory on $Z^{12}$, where the term takes
the form $\int_{Z^{12}} G_4 \wedge I_8$.
Extensions beyond twelve dimensions are considered in~\cite{OP2}.
What we would like to analyze is how this term behaves upon reduction from $Z^{12}$ to its 10-dimensional corner
$M^{10}$.

The one-loop term lifted to twelve dimensions takes the form
\begin{gather}
G_4 \wedge I_8=\left( \tfrac{1}{2}{\lambda} -a \right) \wedge \left(\tfrac{1}{48}\big({p_2 -\lambda^2}\big) \right),
\qquad
\lambda=\tfrac{1}{2}{p_1}.
\label{1-loop}
\end{gather}
We will concentrate at the prime $p=3$, i.e.\
work 3-adically, so that the obstruction to the Fivebrane structure $\frac{1}{6}p_2$, in the sense of~\cite{SSSII}, as
well as the term $\frac{1}{48}{p_2}$ in expression~\eqref{1-loop} both can be ef\/fectively viewed as the fractional class
$\frac{1}{3}{p_2}$.
We will concentrate on the $\lambda \wedge p_2$-term in~\eqref{1-loop}.
Passing to the complex case, taking into account the fact that we are taking $Z^{12}$ to be a~complex manifold, this
part of the action can be written in terms of the Chern classes of $Z^{12}$.
Indeed, if $E$ is a~general complex vector bundle then for the underlying real bundle $E_{\mathbb R}$ one has
\begin{gather*}
p_1(E_{\mathbb R}) \equiv -2c_2(E) \mod c_1(E)
\qquad
\text{and}
\qquad
\big(p_2-\lambda^2\big)(E_{\mathbb R}) \equiv 2 c_4 (E) \mod c_1(E).
\end{gather*}
So if we impose the condition that $E$ is an SU-bundle so that $c_1(E)=0$ (this is satisf\/ied in the physically favorable
setting of the tangent bundle of Calabi--Yau manifolds), then we get for the action
\begin{gather}
\int_{Z^{12}} c_2 \cdot \tfrac{1}{3}c_4.
\label{c2c4}
\end{gather}
This Chern number can be viewed as an analog of the term $\int_{X^{10}}\frac{1}{2}c_2 \cdot c_3$ that appears in the
calculation of the partition function in type IIA via K-theory and $E_8$ gauge theory~\cite{DMW}.
The expression~\eqref{c2c4} is a~12-dimensional analog with the prime 3 taking the place of the prime~2.

Now we reduce expression~\eqref{c2c4} to the corner $M^{10}$.
The general form of the term at the level of dif\/ferential forms is
\begin{gather*}
\int_{M^{10}}\tfrac{1}{3}CS_3 \wedge CS_7,
\end{gather*}
a~product of a~Chern--Simons 3-form and a~Chern--Simons 7-form, both of gravitational type.
We would like to see how this f\/its into the description of the corner via the $f$-invariant (see Section~\ref{Sec f}).
The construction of the $f$-invariant requires a~presentation of the framed ten-manifold $M^{10}$ as a~corner of
codimension two of an almost complex twelve-manifold $Z^{12}$ with suitable splitting of the stable tangent bundle.
We illustrate this with the main physical example, which turns out to also be the main example in the mathematical
construction of the $f$-invariant in~\cite{Lau}.
The group $\text{Sp}(2)$ admits dif\/ferent framings.
Equipped with its left-invariant one, $\text{Sp}(2)$ represents a~generator of $\pi_{10}^sS^0_{(3)}$, i.e.\
the element $\beta$ at the prime 3~\cite{Kn, Lau}.
Note that there is also a~bounding one, as stated previously.
This example, which is a~3-sphere bundle over the 7-sphere, is relevant in string theory in relation to nonrepresentable
cycles~\cite{ES2}.

{\bf Example: $\boldsymbol{M^{10}=\text{Sp}(2)}$.} This example is described from the point of view of K-theory at the prime 3
in~\cite{ES2}.
We will derive conditions on the (lifted) one-loop term arising from a~condition for Sp(2) to be the corner of the
12-manifold $Z^{12}$.
Let $[Z^{12}, \partial Z^{12}]$ be the relative fundamental class.
Pairing cohomology classes with this homology class amounts to detecting those classes which are nontrivial on $Z^{12}$
but become trivial on the boundary $Y^{11}$.
Such a~situation is described in detail in~\cite{S-Sig}.
Now let us consider the condition for the corner to be~Sp(2).
To that end, we will take $Z^{12}$ as a~(U,fr)$^2$-manifold with a~splitting of the stable tangent bundle $T^1Z^{12}
\oplus T^1 Z^{12}$, and consider the corresponding Chern classes $\{ c_1, \dots, c_6\}$ and $\{\tilde{c}_1, \dots,
\tilde{c}_6\}$ of the two subbundles $T^0Z^{12}$ and $T^1 Z^{12}$, respectively.
The condition derived in~\cite{Lau} is given by the $f$-invariant and depends on all possible Chern numbers of total
degree 6; however, if we take $c_1=0=\tilde{c}_1$ (that is complex instead of almost complex as we had before, leading
to~\eqref{c2c4})\footnote{This conditions natural from the point of view of string theory.
Many known compactif\/ications have vanishing f\/irst Chern class.
Furthermore, the bundles associated to the setting in which the one-loop polynomial arises all satisfy this condition.}
then the condition becomes the statement that
\begin{gather*}
\left\langle \tilde{c}_2 \cdot \tfrac{1}{3} c_4 - c_2 \cdot \tfrac{1}{3} \tilde{c}_4, \big[Z^{12}, \partial Z^{12}\big]
\right\rangle \in {\mathbb Q}/{\mathbb Z}_{(3)} \cong {\mathbb Z}_{3^\infty}
\qquad
\text{has~order~} 3.
\end{gather*}
This can be viewed as our condition on the one-loop term of the topological action reduced to the corner, as
in~\eqref{c2c4}.

\subsection{Topological modular forms and Tate K-theory}
\label{Sec tmf}

There are two (related) spectra called Tate K-theory $K_\text{Tate}$; one   is $k[[q]]$ and another is $K\wedge {\rm tmf}$,
with a~map between them which may be interpreted as a~faithfully f\/lat extension in an appropriate category.
Laures identif\/ies the K-theory of elliptic cohomology with Katz's universal ring of divided congruences.
Tate K-theory is already proposed in~\cite{KS1, KS3, S-elliptic,tcu} to essentially describe the elliptic
ref\/inement of the partition function.
The point we make here is that the ring of divided congruences is already present in that theory, and hence is of
physical signif\/icance in the current context.

\looseness=1
{\bf The ring of divided congruence.} The Eisenstein series $E_4$ and $E_6$ with $ E_{2k}=1-\frac{4k}{B_{2k}}
\sum\limits_{n=1}^\infty\Big(\sum\limits_{d\mid n}d^{2k-1}\Big)q^n $ generate the graded ring of modular forms over the complex numbers.
One can capture congruences between modular forms by considering the ring $D$ where all congruences between modular
forms take place.
For example, $E_4 \equiv 1 \mod 240$, corresponds to the class $\tfrac{1}{240}({E_4 -1})\in D$.
However, over the 3-adic integers, one has to consider the ring $D$ of divided congruences: the elements of $D$ are
those 3-adically convergent series in $
\sum f_i$ of (inhomogeneous) modular forms over ${\mathbb Q}_3$ such that the $q$-expansion $\sum f_i (q)= \lambda (\sum
f_i)$ has coef\/f\/i\-cients~$\lambda$ in~${\mathbb Z}_3$.

{\bf Tate K-theory.} Tate K-theory~\cite{AHS, LK1} is def\/ined via the Tate curve, given by $ y^2 + xy=x^3 + a_4x + a_6
$, where the coef\/f\/icients are given by
\begin{gather*}
a_4 = \tfrac{1}{48}(1-E_4) \in {\mathbb Z}[[q]],
\qquad
a_6 = \tfrac{1}{576}(1-E_4) + \tfrac{1}{864}(E_6 -1) \in {\mathbb Z}[[q]].
\end{gather*}
The Tate curve is already def\/ined over the subring $D
\subset {\mathbb Z}[[q]]$ of divided congruences.
Tate K-theory is simply $K[[q]]$ so that the coef\/f\/icient ring is $\pi_0K_\text{Tate}
={\mathbb Z}[[q]]$.
At the level of the moduli space of curves, this theory is obtained by working formally in the neighborhood of the
$j$-invariant taking the value $j=\infty$.
Note that if $q=0$ then $a_4=0=a_6$.
So we see that elements in $D$ correspond to higher order terms beyond the classical term.
We interpret this, in the spirit of~\cite{S-elliptic}, as considering low values of the coupling constant $q$, so that
we view Tate K-theory as a~sort of a~`perturbative elliptic cohomology'.
The sigma-orientation lifts to a~map $MU \langle 6 \rangle \to K[[q]]$ and the invariant $\pi_*\text{MSpin} \to {\mathbb
Z}[[q]]$ asscoiated to the $
\sigma$-prientation on $K_\text{Tate}
$ is the Witten genus~\cite{AHS} $\varphi_W(M)$  $\in {\mathbb Z}[[q]]$.
More on the String condition in string theory can be found in~\cite{String, SSSIII}.

{\bf Modular forms with respect to the congruence subgroup $\boldsymbol{\Gamma=\Gamma_1(3)}$.} Consider the subgroup $\text{SL}(2,
{\mathbb Z}_3)=\text{SL}(2, {\mathbb Z})/\Gamma (3)$ of $\text{SL}(2, {\mathbb Z})$, with the kernel $\Gamma(3)$ a~proper
subgroup of the congruence subgroup
\begin{gather*}
\Gamma_1(3) = \left\{ \left(
\begin{matrix}
1 & *
\\
0 &1
\end{matrix}
\right) \ {\rm mod}\; 3 \right\}
\subset \text{SL}
(2, {\mathbb Z}).
\end{gather*}
The ring of modular forms over $\Gamma=\Gamma_1(3)$ is generated by the two series
\begin{gather*}
E_1 = 1 + 6
\sum\limits_{n=1}
^\infty
\sum\limits_{d\mid n}
\left( \tfrac{d}{3}\right)_{L}q^n, 
\qquad
E_3 = 1 -9
\sum\limits_{n=1}
^\infty
\sum\limits_{d\mid n}
\left( \tfrac{d}{3}\right)_{L}d^2q^n,
\end{gather*}
where $\left( \tfrac{d}{3}\right)_{L}$ denotes the Legendre symbol.
Thus in dealing with the $f$-invariant, in addition to the modular forms $E_4$ and $E_6$, one might encounter the forms
$E_1$ and $E_3$.

\subsection[Chromatic level 2: Heterotic corners and the $f$-invariant]{Chromatic level 2: Heterotic corners and the $\boldsymbol{f}$-invariant}
\label{Sec f}

We have used the $f$-invariant at the level of cohomology classes.
We now show that the ref\/ined version, the geometric $f$-invariant~\cite{BN, Bod}, also captures part of the dynamics and
anomalies of the heterotic corner.

{\bf Elliptic genera and $\boldsymbol{\langle 2\rangle}$-manifolds.} The elliptic genera provide interesting invariants for $\langle
2\rangle$-manifolds.
Instead of ${\mathbb Q}/{\mathbb Z}$ for the $e$-invariant, one wants to consider some values that `combine' ${\mathbb
Q}/{\mathbb Z}$ with modular forms.
For a~$12$-manifold, this is done by Katz's ring of divided congruences~\cite{Lau}.
A framed 10-manifold is the corner of a~(U, fr)$^3$-manifold if and only if the $f$-invariant gives an integral
inhomogeneous modular form for two levels $\geq 2$ which are relatively prime to each other.

Let $X$ be a~closed $U$-manifold.
Then the elliptic genus of $X$ has an integral $q$-expansion~\cite{HBJ}.
Let $M^{10}$ be a~codimension-two corner of a~$(U, \text{fr})$-manifold $Z^{12}$.
The classical constant term and the quantum nonzero $q$ term are given, respectively, by
\begin{gather*}
\text{Ell}_0=\text{Ell}\big|_{q=0},
\qquad
\widetilde{\text{Ell}}=\text{Ell} - \text{Ell}_0.
\end{gather*}
We consider the splitting of the tangent bundle into two bundles $V_1$ and $V_2$.
Using the relative Chern classes of the split tangent bundle, the {\it $f$-invariant} of the framed bordism class of
$M^{10}$ is def\/ined to be~\cite{Bod}
\begin{gather}
f(M^{10})\equiv \langle (\text{Ell}(V_1) -1)(\text{Ell}_0(V_2) - 1), [Z^{12}, \partial Z^{12}] \rangle
\mod \overline{D}_6^\Gamma,
\label{Eq ell}
\end{gather}
where $\overline{D}_6^\Gamma$ is described as follows.
Denote by $M_6^\Gamma$ the graded ring of modular forms with respect to $\Gamma=\Gamma_1(3)$ which expand integrally,
i.e.\
which lie in ${\mathbb Z}^\Gamma[[q]]$.
The ring of divided congruences~$D^\Gamma_6$ consists of those rational combinations of modular forms which expand
integrally.
Then $\overline{D}_6^\Gamma= D_6^\Gamma + M_0^\Gamma \otimes {\mathbb Q} + M_6^\Gamma \otimes {\mathbb Q}$.
Hence $f$ takes values in $\overline{D}_6^\Gamma \otimes {\mathbb Q}/{\mathbb Z}$, and thus is a~natural generalization
of the $e$-invariant, which takes values in ${\mathbb Q}/{\mathbb Z}$~\cite{Lau}.

Note that in~\eqref{Eq ell} one factor in the integrand is ref\/ined while the other factor is classical.
This can be viewed as a~heterotic analog of the elliptic ref\/inement of the one-loop term in type~IIA string theory
in~\cite{String}.
There, only the one-loop polynomial was ref\/ined from $I_8$ to $I_8(q)$, while the $C$-f\/ield and its f\/ield strength $G_4$
remained classical.
The end result is that the whole term $G_4 \wedge I_8$ is ref\/ined to $q$-expansions.

{\bf The geometric $\boldsymbol{f}$-invariant.} We now consider a~connection on the tangent bundle of the $(U, {\rm
fr})^2$-manifold $Z^{12}$ and hence induced connections $\nabla_1$ and $\nabla_2$ on the two bundles $V_1$ and $V_2$.
Consider compatible connections, that is ones which preserve trivializations on the faces, i.e.\
require that they restrict to pure gauge ones on the faces.
Let $M^{10}$ be a~closed ten-manifold which is a~codimension two corner of a~$(U, \text{fr})^2$-manifold $Z^{12}$ of
dimension twelve.
Then $M^{10}$ inherits the splitting of its framing.
Then, using compatible connections, the {\it geometric $f$-invariant} is~\cite{Bod}
\begin{gather*}
\check{f}\big(M^{10}, \nabla_1|_M, \nabla_2|_M\big)\equiv
\left[ \int_{Z^{12}} \widetilde{\text{Ell}}(\nabla_1) \text{Ell}_0(\nabla_2)
\mod \overline{D}_6^\Gamma \right] \in \bigoplus_{k=0}^6 M_k^\Gamma \otimes {\mathbb R},
\end{gather*}
where the l.h.s.\ is def\/ined by considering its residue modulo $\overline{D}_6^\Gamma$.
This expression is congruent to zero mod $\overline{D}_6^\Gamma$ when $Z^{12}$ has an empty corner~\cite{Bod}.
We view this as the topological contribution to the ef\/fective action at the corner.
If $M^{10}$ is the codimension-three corner of a~$(U, \text{fr})^3$-mani\-fold~$W^{13}$ then the $f$-invariant of~$M^{10}$
is trivial.
Therefore, as indicated earlier, in order to detect {\it nontrivial} elliptic cohomology information on~$M^{10}$, the
manifold~$Z^{12}$ itself cannot be a~boundary.

{\bf Products.} We now consider heterotic string theory on product manifolds $M^{10}=X_1^n \times X_2^{10-n}$, viewed as
a~corner of the 12-dimensional manifold $Z^{12}$ and physically interpret the product formulae of~\cite{Bod2}.
As explained in the introduction, for framing as well as from the structure of the physical f\/ields, it is natural to
consider the factors to be odd-dimensional, with the main example being $S^3 \times S^7$, as a~corner of the
twelve-manifold $\mathbb{D}^4 \times \mathbb{D}^8$.
In this case of the general product the $f$-invariant is determined by the complex $e$-invariant of the factors
\begin{gather}
\check{f}(X_1^n \times X_2^{10-n}) \equiv m(X_1^n) e_{\mathbb C}(X_2^{10-n}) \equiv -m(X_2^{10-n}) e_{\mathbb C}(X_1^n),
\label{Eq pro}
\end{gather}
where $m(X_i)$ is any modular form of weight $(\text{dim}X_i +1)/2$ with respect to the group\footnote{Note that the
congruences of~\eqref{Eq ell} carry over to arbitrary levels.} $\Gamma=\Gamma_1(3)$ such that $m(X_i)\equiv e_{\mathbb
C} (X_i)\mod{\mathbb Z}^\Gamma [[q]]$.
We can have modular forms of weight 2 and 4 by taking the factors to be 3- and 7-dimensional, respectively.
This is the correct structure detected in~\cite{String} in the following sense.
The 11-dimensional one-loop term gets ref\/ined only as far as the 8-dimensional one-loop polynomial $I_8$ is concerned
with the $C$-f\/ield part still being classical.
As pointed out in~\cite{String}, one can consider the complimentary point of view where the $C$-f\/ield itself is ref\/ined
while $I_8$ remains classical.
In the decomposition of the ten-dimensional manifold into a~product, we see that we have a~physical manifestation of the
two formulations in formula~\eqref{Eq pro}.
Note that if one of the factors has a~trivial $e$-invariant then the geometric $f$-invariant of the product is congruent
to zero.

{\bf Examples.
1.~The product $\boldsymbol{M^{10}=S^3 \times S^7}$.}  Consider the 3-sphere $S^3$ as the sphere bundle of the Hopf line bundle over
$S^2$.
Framing the base and the vertical tangent gives a~framing for the total space.
The complex $e$-invariant is $e_{\mathbb C}(S^3)= -\frac{1}{12}$, while the real $e$-invariant $e_{\mathbb R}(S^3)$ is
either $-\frac{1}{24}$ or $\frac{11}{24}$, so that $S^3$ represents $\nu$.
Then $m(\nu)=\frac{1}{12}E_1^2$, so that $\overline{m}(\nu):=m(\nu) - e_{\mathbb C}(\nu)= \frac{1}{12}(E_1^2-1)$.
The 7-sphere $S^7$, considered as the sphere of the quaternionic line bundle over $S^4$, represents $
\sigma$.
Then $\mathfrak{e}_{\mathbb C} (\sigma)=\frac{1}
{240}$.
One can take $m(
\sigma)=\frac{1}
{240}E_4$ so that $\overline{m}(
\sigma):= m(\sigma) - e_{\mathbb C}(\sigma)=\frac{1}
{240}(E_4-1)$.
The geometric $f$-invariant for the product is then~\cite{Bod2}
\begin{gather*}
\check{f}\big(S^3 \times S^7\big)\equiv \tfrac{1}{12\cdot 240} (E_4 -1).
\end{gather*}

  {\bf 2.~The product $\boldsymbol{M^{10}=(S^1)^3 \times S^7}$.} For $\eta \in \pi_1^\text{S} \cong Z_2$, $e_{\mathbb C}(\eta)=\frac{1}{2}$.
However, note here that $e_{\mathbb C}(\eta^3)=0$.
Consequently,
\begin{gather*}
\check{f}\big(\big(S^1\big)^3 \times S^7\big) \equiv 0 \cdot \tfrac{1}{240}(E_4 -1) \equiv 0.
\end{gather*}
Other products involving combinations of $S^1\cong\text{U}(1)$ factors with $S^3\cong\text{SU}(2)$ or $S^7$ can be treated
similarly.

The physical interpretation we provide for these examples of~\cite{Bod2} illustrate the reduction to the corner of the
topological term of the M-theory ef\/fective action is captured by a~topological index in twelve dimensions.
Since $E_4=1+ {\cal O}(q)$ then the geometric $f$-invariant detects ${\cal O}(q)$ information.
We hope that this paper helps provide some insight into the role of elliptic cohomology in M-theory, which certainly
deserves a~better understanding and further investigation.
In particular, we hope that the physical setting we provided will help in understanding the $f$-invariant.

{\bf Hierarchy of topological theories.} The proposals and connections put forth here for M-theory can be applied to the
M-branes, i.e.\
to the branes inside M-theory.
Indeed, M-branes with faring and/or corners are studied recently in~\cite{M2M5-framing}, where tantalizing connections
to various topological invariants, including the $f$-invariant, are uncovered.
Furthermore, one can study aspects of both M-branes as well as M-theory itself in the context of extended topological
quantum f\/ield theories (TQFTs).
Such theories require corners of increasing codimension all the way up to the maximal, when the theory is fully
extended.
Our study here concerns, together with~\cite{S-cob, DMW-corner}, point to viewing a~sector in M-theory as a~2-extended
TQFT.
The work~\cite{M2M5-framing} also points to a~similar description for the M-branes themselves.
In all cases one has manifolds of dimension $4k$ with boundary of dimension $4k-1$ and codimenison-2 corner of dimension
$4k-2$, with $k=1, 2, 3$ for the M2-brane, the M5-brane, and M-theory (spacetime) respectively.
The picture that we advocate for the above theories can be captured by the following table

\begin{table}[h!]
\centering
\small
\begin{tabular}{|@{\,\,}c@{\,\,}|@{\,\,}c@{\,\,}|@{\,\,}c@{\,\,}|@{\,\,}c@{\,\,}|}
\hline
Dimension & Type of theory & Boundary/corner structure & Invariants\tsep{2pt}\bsep{2pt}
\\
\hline
\hline
$4k$ & Topological gauge theory & manifold with corners of codim-2 & primary classes\tsep{2pt}\bsep{2pt}
\\
\hline
$4k-1$ & Chern--Simons theory & boundary manifold with boundary & secondary classes\tsep{2pt}\bsep{2pt}
\\
\hline
$4k-2$ & Wess--Zumino--Witten theory & corner of codim-2 & tertiary classes\tsep{2pt}\bsep{2pt}
\\
\hline
\end{tabular}
\end{table}

  The three cases give a~ladder of structures of the form 2-3-4, 6-7-8, and 10-11-12.
The TQFT aspect of this is studied extensively in~\cite{lpqft} within the cobordism hypothesis, with interpretation as
ladders of theories of the form \{Topological gauge theory\}~-- \{Chern--Simons theory\}~-- \{Wess--Zumino--Witten theory\}.
One main point to highlight here is that in order for this to work, one changes the interpretation of the forms at the
middle stage, i.e.\
at the boundary Chern--Simons theory, where the form which is an $n$-connection is now interpreted (appropriately) as an
$n$-curvature, where the value of the positive integer $n$ depends on the specif\/ic theory.

The top-dimensional case is M-theory itself, with a~boundary and a~corner, sitting in the following schematic diagrams
$$
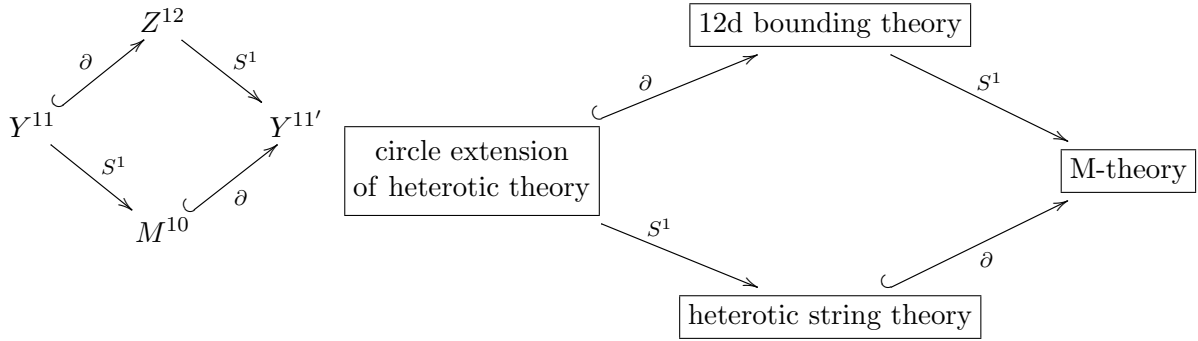
\xymatrix{ & Z^{12} \ar[dr]^{S^1} &
\\
Y^{11} \ar[dr]^{S^1} \ar@{^{(}->}[ur]^\partial && Y^{11'}
\\
& M^{10} \ar@{^{(}->}[ur]_\partial & }
\ \!
\xymatrix{ & \fbox{\text{12d~bounding~theory}} \ar[dr]^{S^1} &
\\
\fbox{$
\begin{array}{@{}c@{}}
\rm circle~extension
\\
\rm of~heterotic~theory
\end{array}
$} \ar[dr]^{S^1} \ar@{^{(}->}[ur]^\partial && \fbox{{\rm M-theory}}
\\
& \fbox{\text{heterotic string theory}} \ar@{^{(}->}[ur]_{\partial} & }
$$
A representative of such a~a situation for the case for the M-branes was studied extensively in~\cite{M2M5-framing}.
The analog in the current setting for the case of M-theory is captured by main example of the product of two closed
disks, leading to a~two-step reduction of the form
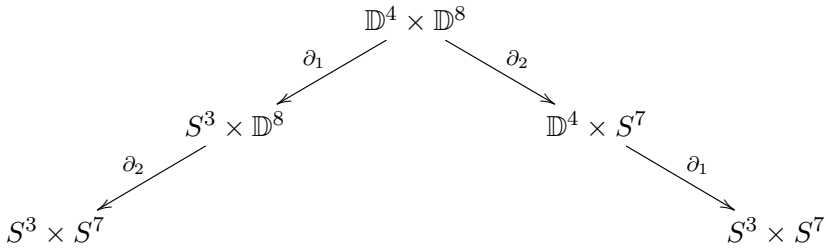
\begin{gather*}
\xymatrix{ && {\mathbb D}^4 \times {\mathbb D}^8 \ar[dl]_{\partial_1} \ar[dr]^{\partial_2} &&
\\
& S^3 \times {\mathbb D}^8 \ar[dl]_{\partial_2}  && {\mathbb D}^4 \times S^7  \ar[dr]^{\partial_1} &
\\
S^3 \times S^7 & &  & & S^3 \times S^7
}
\end{gather*}
We now can describe this from the point of view of cobordism.
We start with the corner $S^3 \times S^7$ and consider it as a~boundary in two dif\/ferent ways, namely for ${\mathbb D}^4
\times S^7$ and for $S^3 \times {\mathbb D}^8$.
These two are related by a~surgery which is implemented by the corresponding space ${\mathbb D}^4 \times {\mathbb D}^8$.
The situation is depicted in this diagram
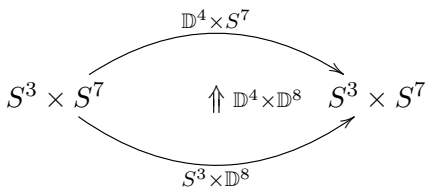
\begin{gather*}
\xymatrix@C+54pt{S^3 \times S^7\rtwocell<10>^{{\mathbb D}^4 \times S^7}_{S^3\times {\mathbb D}^8}{^{\hspace*{18mm}{\mathbb D}^4 \times {\mathbb D}^8}}& S^3 \times S^7}
\end{gather*}
which describes a~2-category of cobordisms, as appropriate in the existence of codimension-2 corners.
This deserves more discussion and will be fully developed elsewhere.

{\bf The (generalized) WZW theory.} The ten-dimensional string theory can be viewed in two ways: First, as
a~codimension-2 corner and, second, as a~(generalized) Wess--Zumino--Witten (WZW) theory.
The f\/irst is emphasized in this paper and the second is highlighted in previous works~\cite{targets,S-gerbe, String,
DMW-boundary}.
Elliptic cohomology on the one hand is used to def\/ined the $f$-invariant, which comes up as the elliptic genus of
codimension-2 manifolds~\cite{Lau}.
Elliptic cohomology is also used to describe the partition function in the heterotic theory; see~\cite{String} for
a~description in this context.
This can also be described via the notion of f\/ibered WZW theory, string-theoretically realizing elliptic
genera~\cite{DS}.
In addition to this general conceptual reasoning, we have provided above an explicit relation to the main topological
term in the theory, namely the one-loop term, also complementing a~similar discussion in~\cite{String}.

\subsection*{Acknowledgements}

The author thanks Ulrich Bunke for explaining his work and Niranjan Ramachandran for discussions on divided congruences.
This research is supported by NSF Grant PHY-1102218.
The author is indebted to the anonymous referees for many corrections and helpful suggestions.

\pdfbookmark[1]{References}{ref}
\LastPageEnding

\end{document}